\documentclass[floatfix,reprint, amsmath,amssymb,aps,superscriptaddress]{revtex4-2}

\usepackage[utf8]{inputenc}

\usepackage{graphicx}
\usepackage{hyperref}
\usepackage{subcaption}
\usepackage{dcolumn}
\usepackage{bm}
\usepackage{physics}
\usepackage[english]{babel}

\begin{document}
\title{Dynamics of two interacting dipolar two-level systems in a multi-mode electromagnetic cavity: sudden death and revival of the entanglement within the Born-Markov approximation}

\author{Loann Quien}
\affiliation{École Normale Supérieure Paris-Saclay, 4 Av. des Sciences, 91190 Gif-sur-Yvette, France}
\affiliation{Institute of Industrial Science, The University of Tokyo, 5-1-5 Kashiwanoha, Kashiwa, Chiba 270-0139, Japan}
\author{Naomichi Hatano}
\affiliation{Institute of Industrial Science, The University of Tokyo, 5-1-5 Kashiwanoha, Kashiwa, Chiba 270-0139, Japan}

\date{\today}

\begin{abstract}
  Interacting dipolar two-level systems form a special class of qubits that interact with a cavity in a particular way. We first prove that the Markovian dynamics of one 1/2-spin in interaction with a quantised magnetic field from a multi-mode cavity at thermal equilibrium is equivalent to a two-level atom interacting in the dipole approximation with the electric field of the cavity. We then use the Born-Markov approximation to study the dynamics of two spins interacting through the antiferromagnetic Heisenberg coupling in the same environment. By solving the GKSL equation, we find the exact expression of the density matrix of the system, with the off-diagonal coherence decay time and spin relaxation time. The concurrence for the stationary state is explicitly derived for any kind of initial state and the role of the singlet state is brought to light. The temporal evolution of the concurrence is numerically computed for different initial states, the phenomenon of sudden death and revival of the entanglement is observed for this dynamics. A detailed analysis of the sudden death and revival of the concurrence is conducted for Werner states, with new analytical results obtained thanks to the solution of the GKSL equation. We finally derive the equations and the stationary concurrence for the XXZ coupling.    
\end{abstract}
\maketitle


\section{Introduction}
The spontaneous emission rate for the dipole interaction with a quantised electric field has been explicitly derived \cite{cohentannoudji1998atom} and experimentally measured and controlled \cite{Purcell1995SpontaneousEP} \cite{noda_spontaneous-emission_2007}. However, the spontaneous emission rate for the Zeeman interaction between a 1/2-spin and a quantised magnetic field is often treated by analogy with the dipole one \cite{https://doi.org/10.1002/lpor.201600268} and rarely rigorously proven. One proof from B. Fain \cite{PhysRevA.37.546} involves the Zwanzig formalism and the derivation of the Chapman-Kolmogorov jump master equation for the probabilities of transition. 

We propose here an alternative proof with another microscopic method from Breuer and Petruccione \cite{breuer_theory_2009}. With the Born-Markov approximation, this proof naturally leads to the derivation of the Gorini–Kossakowski–Sudarshan–Lindblad (GKSL) master equation \cite{gorini_completely_1976} \cite{lindblad_generators_1976} for this spin system and justify the expected equivalence with a two-level atom in the dipole interaction. We also discuss the value of the magnetic decay time obtained with this model by comparing it to the standard values of decay times for electric dipoles. 

Then, we use this result to study the dynamics of two interacting two-level systems (spins or two-level atoms) in a thermal cavity and microscopically derive the corresponding GKSL master equation. Li and Xu studied the steady states of this model by setting the initial condition as Werner states \cite{li_stationary_2008}, while Wu \textit{et al.} realised an exact numerical study of a similar model without the Born-Markov approximation \cite{wu_exact_2013}. Our study uses a different model from the previous ones by considering the $\vec{S}\cdot \vec{B}$ coupling instead of the $S_x B_x$ or $S_z B_z$. 

We provide analytical and numerical results for the dynamics, such as the relaxation time, the decoherence time and the study of the entanglement between the two systems. We find that this simple approach within the Born-Markov approximation leads the concurrence to the sudden death and revival behaviour, which was only observed, in the Markovian case, for two atoms following the Lehmberg–Agarwal  master equation in a vacuum field \cite{PhysRevA.74.024304} or for non-interacting systems in a common environment by using a non-perturbative approach \cite{PhysRevA.79.042302}. We use the results for the dynamics to derive some new analytical properties for the sudden death and revival. Our proof shows that these results hold for any kind of interacting two-level systems in an SU(2)-like dipole interaction with the electromagnetic field of a cavity.

\section{One spin dynamics}
\subsection{Model}
Consider $\vec{B_0} = B_0 \hat{z}$ a classical magnetic field and $\vec{B}$ a quantised magnetic field from a cavity, in the Schrödinger picture:
\begin{equation}
 \vec{B}=i \sum_{\vec{k}, \lambda}\sqrt{\frac{ \hbar}{2\epsilon_0\omega_kV}}(\vec{k} \times \vec{e}_{\vec{k}\lambda})(b_\lambda(\vec{k})-b^\dagger_\lambda(\vec{k})), 
 \label{B}
\end{equation}
where $V$ is the quantisation volume of the cavity, $\vec{e}_{\vec{k}\lambda}$ the polarisation and $\omega_k=|\vec{k}|c$. We consider a 1/2-spin interacting through the Zeeman effect with both magnetic fields. 

The interaction with the classical field gives rise to a two-level system. Under the basis $\{\ket\downarrow,\ket\uparrow \}$, the Hamiltonian $H_S$ of the two-level system is
\begin{equation}
H_S=g\mu_B B_0 S_z =\hbar \omega_0 S_z,
\end{equation}
where $S_z=(\ket\uparrow \bra\uparrow - \ket\downarrow \bra\downarrow)/2$. On the other hand, the interaction between the system and the quantised cavity is described by the perturbative Hamiltonian 
\begin{equation}
H_I=g\mu_B \vec{S}\cdot \vec{B},
\end{equation}
with the 1/2-spin operator $\vec{S}=\vec{\sigma}/2$ and $\{\sigma_x,\sigma_y,\sigma_z\}$ the Pauli matrices. 
The normal-ordered Hamiltonian $H_B$ of the cavity is
\begin{equation}
\label{Hb}
    H_B=\sum_{\vec{k}, \lambda}\hbar \omega_k b^\dagger_\lambda(\vec{k})b_\lambda(\vec{k}).
\end{equation}

To complete the framework of the study, we use the weak-coupling Born-Markov approximation \cite{breuer_theory_2009}, which means the following:
\begin{itemize}
    \item We consider a first-order perturbation in $H_I$. 
    \item The total density matrix $\rho(t)$ can always be written as the tensor product of the density matrix of the system $\rho_S(t)$ and  of the density matrix of the cavity $\rho_B$, as in $\rho(t)=\rho_S (t) \otimes \rho_B$.
    \item The dynamic of the system is Markovian, and thus the density matrix of the system follows the Gorini–Kossakowski–Sudarshan–Lindblad (GKSL) master equation \cite{gorini_completely_1976} \cite{lindblad_generators_1976}.
\end{itemize}

The first two points, known as the Born approximation, are relevant when the quantised magnetic field of the cavity is small regarding the classical field, and when the environment is much larger than the system. Hence, as long as the number of photons in the cavity does not become macroscopic and by considering a cavity with an infinite number of degrees of freedom, the Born approximation is relevant. The Markov approximation will be discussed later in Sec. \ref{proof_gksl} .

Moreover, we consider the cavity at thermal equilibrium with an exterior heat bath $\beta$:
\begin{equation}
    \label{therm}
    \rho_B=\frac{\exp(-\beta H_B)}{\mathrm{tr}[\exp(-\beta H_B)]}.
\end{equation}
In order to have a non-macroscopic number of photons in the cavity the temperature cannot be arbitrarily high.  

This set-up is almost the same as the one for a two-level system interacting in the dipole approximation with a quantised electric field at thermal equilibrium. In fact, we will show that the resulting dynamics is the same, with an analogous spontaneous emission rate.

\subsection{Calculation}

\subsubsection{Jump operators of the GKSL equation}
We first decompose the spin operator $\vec{S}$ into eigenoperators of $H_S$. Let $\Pi(\epsilon)$ denote the projector on the $H_S$ eigenspace corresponding to the eigenvalue $\epsilon$. For $\epsilon=\hbar \omega_0/2$, for instance,  we have $\Pi(\epsilon)=\ket{\uparrow}\bra{\uparrow}$. We then obtain 
\begin{equation}
\label{deco}
  \vec{A}(\omega):=g\mu_B\sum_{\epsilon'-\epsilon=\hbar\omega}\Pi(\epsilon)\vec{S}\Pi(\epsilon'), 
\end{equation}
and we can write $H_I=\sum_{\omega} \vec{A}(\omega)\cdot \vec{B}$. We only have one transition and hence we find $\vec{A}(\omega_0)$ and $\vec{A}(-\omega_0)$ such as
\begin{equation}
 \vec{A}(\omega_0)=\vec{A}(-\omega_0)^\dagger=g\mu_B\mel{\downarrow}{\vec{S}}{\uparrow}\ket{\downarrow}\bra{\uparrow}.  
\end{equation}
These operators are the jump operators used in the GKSL equation.

\subsubsection{Spectral correlation tensor}

The next step is the calculation of the spectral correlation tensor \cite{breuer_theory_2009}:
\begin{equation}
\label{gentens}
  \Gamma_{ij}(\omega,t)=\frac{1}{\hbar^2}\int_0^\infty dse^{i\omega s}\expval{B_i(t)B_j(t-s)}, 
\end{equation}
where $\vec{B}(t)=e^{\frac{i}{\hbar}H_B t}\vec{B}e^{-\frac{i}{\hbar}H_B t}$ is the representation in the interaction picture of the quantised magnetic field defined in Eq. \eqref{B} and $\expval{B_i(t)B_j(t-s)}=\mathrm{tr_B}[B_i(t)B_j(t-s) \rho_B]$, with tr$_B$ denoting the trace operation over the cavity degrees of freedom.
By plugging the expression of $\vec{B}(t)$ into Eq. \eqref{gentens} we find in general 
\begin{equation}
\begin{aligned}
\Gamma_{ij}(\omega,t)=\sum_{\vec{k}, \vec{k'},\lambda,\lambda'}\frac{\hbar}{2\epsilon_0V\sqrt{\omega_k \omega_{k'}}}(\vec{k} \times \vec{e}_{\vec{k}\lambda})_i (\vec{k}' \times \vec{e}_{\vec{k}'\lambda'})_j \\
\times \frac{1}{\hbar^2} \int_0^\infty ds \expval{b_\lambda(\vec{k})b^\dagger_{\lambda'}(\vec{k}')} e^{i(\omega_{k'}-\omega_k)t - i(\omega_{k'}-\omega)s}
\\
+\expval{b^\dagger_\lambda(\vec{k})b_{\lambda'}(\vec{k}')} e^{-i(\omega_{k'}-\omega_k)t + i(\omega_{k'}+\omega)s}
\\
-\expval{b_\lambda(\vec{k})b_{\lambda'}(\vec{k}')} e^{-i(\omega_{k'}+\omega_k)t + i(\omega_{k'}+\omega)s}
\\
-\expval{b^\dagger_\lambda(\vec{k})b^\dagger_{\lambda'}(\vec{k}')} e^{i(\omega_{k'}+\omega_k)t - i(\omega_{k'}-\omega)s}.
\end{aligned}
\end{equation}

We then use the thermal equilibrium state defined in Eq. \eqref{therm} for $\rho_B$. This state is stationary, and therefore the correlations functions of the reservoir are homogeneous in time:
\begin{equation}
  \expval{B_i(t)B_j(t-s)}=\expval{B_i(s)B_j(0)},
\end{equation}
and hence the correlation tensor \eqref{gentens} is independent of time. We use the expectation values
\begin{gather}
    \expval{b_\lambda(\vec{k})b^\dagger_{\lambda'}(\vec{k}')} = \delta_{kk'}\delta_{\lambda \lambda'}(1+N(\omega_k)), \\
    \expval{b^\dagger_\lambda(\vec{k})b_{\lambda'}(\vec{k}')} = \delta_{kk'}\delta_{\lambda \lambda'} N(\omega_k), \\
     \expval{b_\lambda(\vec{k})b_{\lambda'}(\vec{k}')}=\expval{b^\dagger_\lambda(\vec{k})b^\dagger_{\lambda'}(\vec{k}')}=0,
\end{gather}
where 
\begin{equation}
    N(\omega_k)=\frac{1}{e^{\beta \hbar \omega_k}-1},
\end{equation}
which yields a spectral correlation tensor \eqref{gentens} in the form
\begin{equation}
\begin{aligned}
\Gamma_{ij}(\omega)=\frac{1}{\hbar^2} \sum_{\vec{k}} \sum_{\lambda}\frac{\hbar}{2\epsilon_0V}\frac{1}{\omega_k}(\vec{k} \times \vec{e}_{\vec{k}\lambda})_i (\vec{k} \times \vec{e}_{\vec{k}\lambda})_j \\
\times \int_0^\infty ds \left[(1+N(\omega_k)) e^{-i(\omega_{k}-\omega)s}
+N(\omega_k) e^{i(\omega_k+\omega)s}\right].
\end{aligned}
\end{equation}

Next we take the continuum limit:
\begin{equation}
\frac{1}{V}\sum_{\vec{k}} \longrightarrow \int \frac{d^3k}{(2\pi)^3}=\frac{1}{(2\pi c)^3}\int_0^\infty d\omega_k \omega_k^2 \int d\Omega.
\end{equation}
This is where the difference with the dipole interaction emerges. We have the term 
\begin{equation}
    \sum_\lambda \int d\Omega (\vec{k} \times \vec{e}_{\vec{k}\lambda})_i (\vec{k} \times \vec{e}_{\vec{k}\lambda})_j
\end{equation}
instead of the easier term for an electric dipole \cite{breuer_theory_2009}
\begin{equation}
\int d\Omega \sum_\lambda(\vec{e}_{\vec{k}\lambda})_i (\vec{e}_{\vec{k}\lambda})_j = \int d\Omega \left( \delta_{ij} - \frac{k_ik_j}{k^2}\right)=\frac{8\pi}{3} \delta_{ij}.
\end{equation}

Nonetheless, we can do the exact calculation. We use the implicit summation over consecutive indices (including $\lambda$ but not $k$) and the Levi-Civita tensor $\epsilon_{ilm}$:
\begin{equation}
(\vec{k} \times \vec{e}_{\vec{k}\lambda})_i (\vec{k} \times \vec{e}_{\vec{k}\lambda})_j= \epsilon_{ilm}\epsilon_{jnp}k_lk_ne^m_{k\lambda}e^p_{k\lambda}.
\end{equation}
We have $e^m_{k\lambda}e^p_{k\lambda}=\delta_{mp}-k_mk_p/k^2$, and hence, by using the relation
\begin{equation}
\epsilon_{ilm}\epsilon_{jnm}=\delta_{nl}\delta_{ij}-\delta_{ni}\delta_{jl}
\end{equation}
and the fact
\begin{equation}
\epsilon_{ilm}\epsilon_{jnp}k_lk_nk_mk_p= (\vec{k}\times\vec{k})_i (\vec{k}\times\vec{k})_j = 0,
\end{equation}
we obtain
\begin{equation}
(\vec{k} \times \vec{e}_{\vec{k}\lambda})_i (\vec{k} \times \vec{e}_{\vec{k}\lambda})_j= k^2 \left(\delta_{ij}-\frac{k_ik_j}{k^2} \right),
\end{equation}
which is followed by 
\begin{equation}
\int d\Omega (\vec{k} \times \vec{e}_{\vec{k}\lambda})_i (\vec{k} \times \vec{e}_{\vec{k}\lambda})_j = \omega_k^2 \frac{8\pi}{3} \delta_{ij} \frac{1}{c^2}.
\end{equation}
The spectral correlation tensor is thus given by:
\begin{equation}
\label{cor tensor}
\begin{aligned}
\Gamma_{ij}(\omega)&= \delta_{ij}\frac{1}{6\pi^2 \epsilon_0 \hbar c^3}\frac{1}{c^2}\int_0^\infty d\omega_k \omega_k^3  \int_0^\infty ds \\ &\left[(1+N(\omega_k)) e^{-i(\omega_{k}-\omega)s}+N(\omega_k) e^{i(\omega_k+\omega)s}\right].
\end{aligned}
\end{equation}
The only difference with the dipole interaction \cite{breuer_theory_2009} is the factor $1/c^2$.

\subsubsection{Master equation}
\label{me}
The calculations are now the same as for a two-level system in the dipole approximation. In the interaction picture, we have the master equation \cite{breuer_theory_2009}
\begin{equation}
\begin{aligned}
\dot{\rho_S}&=\sum_{\omega,\omega'}\sum_{i,j}e^{i(\omega'-\omega)t}\Gamma_{ij}(\omega) \\
&\left(A_j(\omega)\rho_SA_i^\dagger(\omega')-A_i^\dagger(\omega')A_j(\omega)\rho_S\right) + \text{h.c.},
\end{aligned}
\end{equation}
 with the sums over $\omega$ and $\omega'$ taking both the transition at $\omega$ and the transition at $-\omega$. By performing the rotating-wave approximation, neglecting the Lamb shift, and writing $\gamma_{ij}(\omega)=\Gamma_{ij}(\omega) + \Gamma_{ji}^*(\omega)$, we obtain
 \begin{equation}
 \begin{aligned}
 \dot{\rho_S}&=\sum_{\omega}\sum_{i,j}\gamma_{ij}(\omega)\left(A_j(\omega)\rho_SA_i^\dagger(\omega)
 -\frac{1}{2}\left\{A_i^\dagger(\omega)A_j(\omega),\rho_S\right\}\right)\\
 &= \gamma_0(1+N(\omega_0))\left(S_-\rho_SS_+-\frac{1}{2}\left\{S_+S_-,\rho_S\right\}\right) + \\ &\gamma_0N(\omega_0)\left(S_+\rho_SS_--\frac{1}{2}\left\{S_-S_+,\rho_S\right\}\right), 
\end{aligned}
\end{equation}
with the spin ladder operators $S_+=\ket{\uparrow}\bra{\downarrow}$ and $S_-=\ket{\downarrow}\bra{\uparrow}$ and $\gamma_0$ the spontaneous emission rate:
\begin{equation}
\label{rate}
\begin{aligned}
\gamma_0&=\frac{\omega_0^3g^2\mu_B^2}{6\pi \epsilon_0 \hbar c^5 }=\frac{B_0^3g^5\mu_B^5}{6\pi \epsilon_0\hbar^4 c^5}=B_0^3\frac{g^5e^5\hbar}{192\pi \epsilon_0m_e^5c^5}\\
&=\frac{g^5}{48}\alpha\left(\frac{eB_0}{m_e}\right)^3 \left(\frac{\hbar}{m_ec}\right)^2\frac{1}{c^2}\\
&=\frac{g^5}{48}\alpha\frac{\omega_c^3\Lambda_c^2}{c^2},
\end{aligned}
\end{equation}
with $\alpha$ the fine-structure constant, $\omega_c$ the cyclotron pulsation of the electron in $B_0$, and $\Lambda_c$ the Compton length of the electron. We see that it is the same as the dipole interaction except that the dipole term $|\vec{d}|^2$, where $\vec{d}$ is the dipole of the system, is replaced by $g^2\mu_B^2/2c^2$. For an usual experimental set-up such as $g=2$ and $B_0=0.1$ T we have $\gamma_0=1.7\cdot 10^{-12} $ s$^{-1}$, which is of the same order as in Ref. \cite{bienfait_controlling_2016}. By writing $\gamma=\gamma_0(2N(\omega_0)+1)$, $\rho_S=(\rho_{ij})_{ij} \in M_2(\mathbf{C})$ in the $\{\ket{\uparrow},\ket{\downarrow}\}$ basis and initially having only the state $\ket{\downarrow}$ populated, we obtain
\begin{gather}
\rho_{12}=\rho_{12}(0)e^{-\frac{\gamma}{2}t},\\
\rho_{11}= \frac{N(\omega_0)}{2N(\omega_0)+1}(1-e^{-\gamma t}),\\
\rho_{22}=1-\rho_{11}.
\end{gather}

\section{Two interacting spins dynamics}

\subsection{Model}
Let us next consider two 1/2-spins in the same environment and the same classical magnetic field $\vec{B}_0$. We let these spins interact through the antiferromagnetic Heisenberg exchange interaction, which results in the following Hamiltonian
\begin{equation}
\label{xxx}
   H_S=\hbar\omega_0(S_1^z+S_2^z) + J\vec{S}_1\cdot\vec{S}_2,
\end{equation}
where $J>0$ is the exchange constant. The interaction Hamiltonian is modified in the same way:
\begin{equation}
\label{xxi}
H_I=g\mu_B(\vec{S}_1+\vec{S}_2)\cdot\vec{B}=g\mu_B\vec{S}\cdot\vec{B}.
\end{equation}
The cavity Hamiltonian stays the same as in Eq. \eqref{Hb}.

\subsection{Calculation}
In this subsection, we microscopically derive the GKSL equation for the system described by Eq. \eqref{xxx}--\eqref{xxi} and Eq. \eqref{Hb}. We briefly discuss the experimental relevance of this model, and then we analytically solve the GKSL equation and study the stationary state and the stationary entanglement.

\subsubsection{Derivation of the GKSL master equation}
\label{proof_gksl}
We first diagonalise $H_S$, using the triplet and singlet basis $\ket{\uparrow\uparrow}, \ket{\uparrow\downarrow^{\pm}} := \frac{1}{\sqrt2}(\ket{\uparrow\downarrow}\pm\ket{\downarrow\uparrow}) \textrm{ and } \ket{\downarrow\downarrow}$ of the system of two 1/2-spins. The eigenvalues of $H_S$ are $\{\hbar\omega_0+J/4,J/4,J/4-\hbar\omega_0,-3J/4\}$ associated to the eigenbasis $\{\ket{\uparrow\uparrow},\ket{\uparrow\downarrow^+},\ket{\downarrow\downarrow},\ket{\uparrow\downarrow^-}\}$, with the classical magnetic field $\vec{B}_0$ breaking the degeneracy of the triplet states. 

The next step is to calculate the jump operators. We have five transition frequencies $\omega_0,2\omega_0,\omega_0\pm J/\hbar,J/\hbar$, with $\omega_0$ corresponding to the two transitions $\ket{\uparrow\uparrow} \longrightarrow\ket{\uparrow\downarrow^+}$ and $\ket{\uparrow\downarrow^+} \longrightarrow\ket{\downarrow\downarrow}$. For each transition, we calculate the matrix element $\mel{\textrm{initial state}}{\vec{S}}{\textrm{final state}}$. We find
\begin{equation}
\begin{aligned}
    \mel{\uparrow\uparrow}{\vec{S}}{\downarrow\downarrow}&=\mel{\uparrow\uparrow}{\vec{S}}{\uparrow\downarrow^-}=\mel{\uparrow\downarrow^+}{\vec{S}}{\uparrow\downarrow^-}=\mel{\downarrow\downarrow}{\vec{S}}{\uparrow\downarrow^-}\\
    &=0, 
\end{aligned}
\end{equation}
\begin{equation}
    \mel{\uparrow\uparrow}{\vec{S}}{\uparrow\downarrow^+}=\frac{1}{\sqrt2}(\hat{x}-i\hat{y})=\mel{\uparrow\downarrow^+}{\vec{S}}{\downarrow\downarrow}    
\end{equation}
which is in accordance with the selection rules for a dipolar transition. By using the definition of Eq. \eqref{deco}, we therefore obtain
\begin{equation}
\vec{A}(\omega_0)=\frac{g\mu_B}{\sqrt2}(\hat{x}+i\hat{y})\left(\ket{\uparrow\downarrow^+}\bra{\uparrow\uparrow}+\ket{\downarrow\downarrow}\bra{\uparrow\downarrow^+}\right).
\end{equation}
The spectral correlation tensor is the same as in Eq. \eqref{cor tensor} because the environment is still described by Eq. \eqref{Hb}. 

The GKSL master equation in the interaction picture is written as
\begin{equation}
\label{gksl}
\begin{aligned}
  \dot{\rho_S} = \gamma_0^{(2)}(1+N(\omega_0))\left(S_-\rho_SS_+-\frac{1}{2}\left\{S_+S_-,\rho_S\right\}\right)\\ + \gamma_0^{(2)}N(\omega_0)\left(S_+\rho_SS_--\frac{1}{2}\left\{S_-S_+,\rho_S\right\}\right),
\end{aligned}
\end{equation}
where $\gamma_0^{(2)}=g^5\alpha\omega_c^3\Lambda_c^2/24c^2$ is twice the spontaneous emission rate of the precedent one-spin system in Eq. \eqref{rate} and
\begin{equation}
   S_+= \begin{pmatrix}
0 & 1 & 0&0\\
0 & 0&1&0 \\
0&0&0&0\\
0&0&0&0
\end{pmatrix}= S_-^\top.
\end{equation}
We see that the triplet behaves as a three-level system only allowed to jump from one level to the the next, without skipping a step. 

We thus proved that, in an electromagnetic cavity, the dynamics of two interacting 1/2 spins coupled to a classical magnetic field is equivalent to the dynamics of two two-level atoms or molecules coupled to each other via the dipole-dipole Heisenberg-like  interaction. What follows is thus theoretically relevant for both systems, but the XXX coupling is not usual for electric dipoles and is experimentally hard to realise while the XXZ coupling has been argued to be easier to materialise \cite{muller_quantum_nodate}.

Experimentally, these results are not directly applicable. One needs to perform additional manipulations to obtain this dynamics. Indeed, we showed that the spontaneous emission rate of the magnetic system is of the order of $10^{-12} $ s$^{-1}$ for usual experimental set-ups. \textit{A priori}, this gives us exceptionally long relaxation times for spin-qubits in a cavity. In comparison, the usual spontaneous emission rate for an electric dipole is of the order of $10^{8} $ s$^{-1}$ \cite{cohentannoudji1998atom}. This means that magnetic dipoles are apparently better candidates for qubits engineering than electric dipoles. 

Note that in practice, spin relaxation is not governed by the pure spontaneous emission. In NV centers for instance, the dominating interaction and cause of relaxation comes from the spin bath formed by the surrounding spins of the carbon nuclei \cite{balasubramanian_ultralong_2009} \cite{amsuss_cavity_2011}. To make the spontaneous emission relevant again, one can enhance it with the Purcell effect \cite{Purcell1995SpontaneousEP} \cite{bienfait_controlling_2016}, which requires to modify the electromagnetic cavity. 

Finally, we discuss the validity of the Markov approximation and rotating-wave approximation. The Markov approximation is valid if the characteristic evolution time of the environment $\tau_B$ is much shorter than the relaxation time of the system $\tau_R$, which means that the environment instantly responds to the sub-system, without storing memory \cite{breuer_theory_2009}. In our case, we have
\begin{gather}
    \tau_B \sim \frac{h}{k_BT} = \frac{4.8 \cdot 10^{-11}}{T [K]} s, \\ \tau_R \sim \frac{1}{\gamma_0^{(2)}} \approx 10^{12} s,
\end{gather}
which means that the Markov approximation for magnetic systems is relevant even for very low temperatures $T\gg 10^{-23}$ K. For electric dipoles systems, we would have $T\gg 10^{-3}$ K.
The rotating-wave approximation is valid if the intrinsic evolution time of the system is much longer than its relaxation time \cite{breuer_theory_2009}:
\begin{equation}
    \gamma_0^{(2)} \ll \omega_0,
\end{equation}
which is true for magnetic and electric systems, because $\omega_0 \approx 10^{10}$ rad$\cdot$s$^{-1}$.

\subsubsection{Dynamics of the diagonal terms}
\label{dynamics}
\begin{figure}[htbp]
    \centering
    \includegraphics[width=\linewidth]{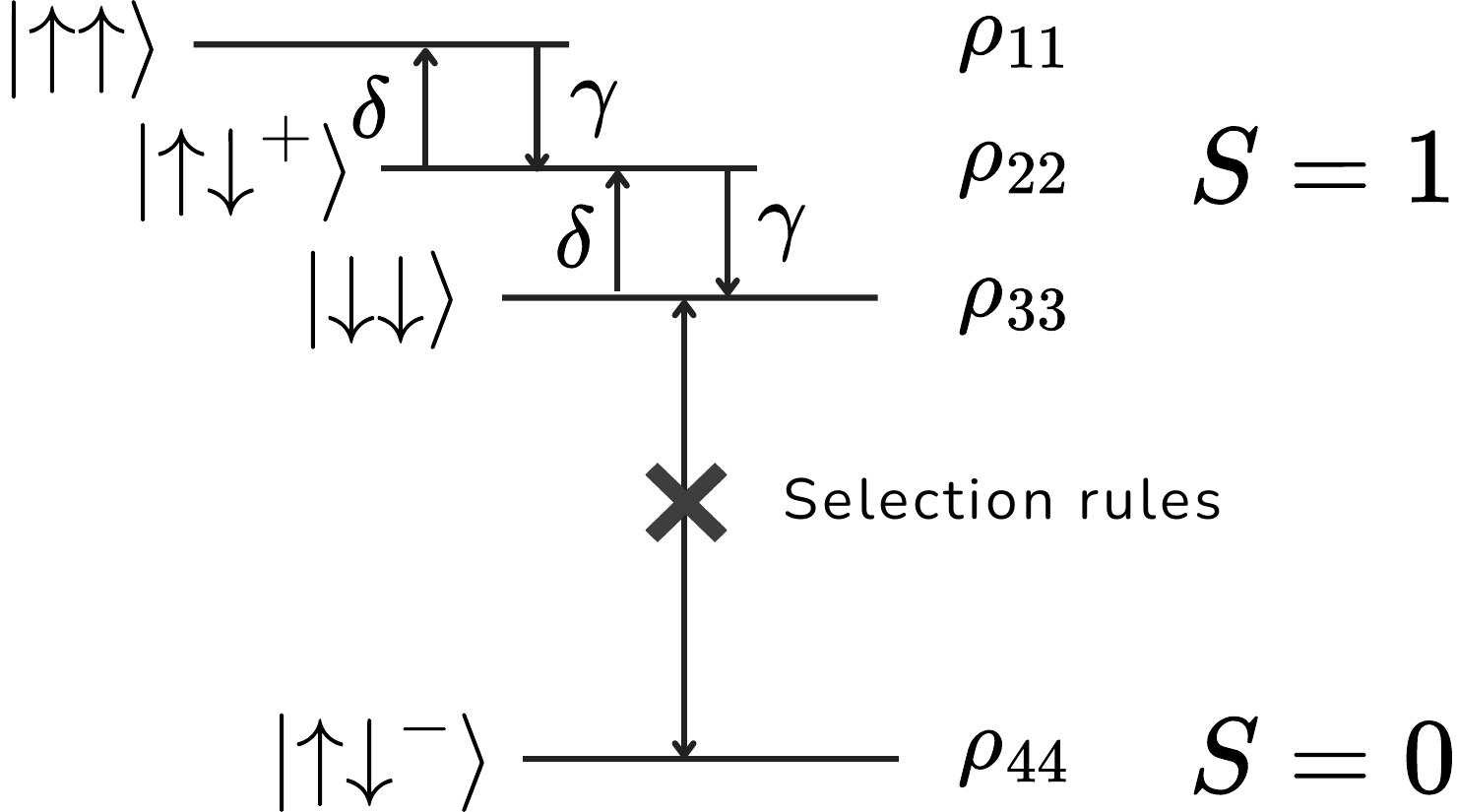}
    \caption{Energy levels of the system and transitions.}
    \label{schema}
\end{figure}

We write $N_0:=N(\omega_0)$, $\gamma:=\gamma_0^{(2)}(1+N_0)$, $\delta:=\gamma_0^{(2)}N_0$ and $\rho_S=(\rho_{ij})_{ij} \in M_4(\mathbf{C})$ in the eigenbasis  $\{\ket{\uparrow\uparrow},\ket{\uparrow\downarrow^+},\ket{\downarrow\downarrow},\ket{\uparrow\downarrow^-}\}$ of $H_S$. We obtain the following system of differential equations for the diagonal terms, \textit{i.e} the population of the different levels:
\begin{gather}
\label{densities1}
    \dot{\rho_{11}}=\delta\rho_{22} - \gamma\rho_{11},\hspace{1cm} \dot{\rho_{33}}=\gamma\rho_{22} - \delta\rho_{33}, \\
    \dot{\rho_{22}}=\gamma(\rho_{11}-\rho_{22}) + \delta(\rho_{33}-\rho_{22}), \\ \dot{\rho_{44}}=0  \label{densities2}.
\end{gather}
We can see that $\gamma$ and $\delta$ represent the decreasing and increasing thermal rates, which naturally leads to a jump master equation. We summarise the situation in Fig. \ref{schema}.

We see that, according to the selection rules, the population of the singlet state is fixed. Because of this, the stationary state of the system is not the thermal equilibrium state ; only the three levels of the triplet are at thermal equilibrium. The stationary solutions of Eqs. \eqref{densities1}--\eqref{densities2} are given by
\begin{gather}
    \rho^s_{22}=\frac{N_0(1+N_0)}{1+3N_0(N_0+1)}(1-\rho_{44}),\\
    \rho^s_{11}= \frac{\delta}{\gamma}\rho^s_{22}=\frac{N_0^2}{1+3N_0(N_0+1)}(1-\rho_{44}),\\
    \rho^s_{33}= \frac{\gamma}{\delta}\rho^s_{22}=\frac{(1+N_0)^2}{1+3N_0(N_0+1)}(1-\rho_{44}),
\end{gather}
which confirms that at high temperatures ($N_0\gg1$) the three levels are equally occupied, while at low temperatures ($N_0\ll1$) $\rho^s_{22} \propto N_0$ and $\rho^s_{11} \propto N_0^2$.

\begin{figure}[htbp]
    \centering
    \includegraphics[width=\linewidth]{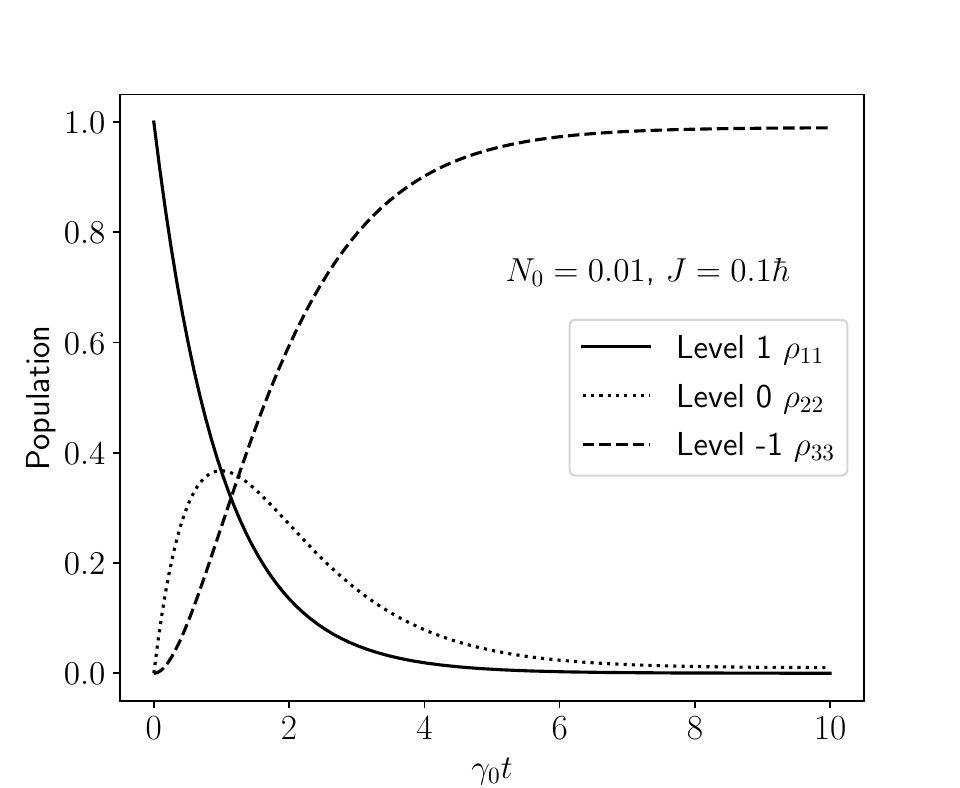}
    \caption{Dynamics of the diagonal terms of the triplet states.}
    \label{fig:density}
\end{figure}

We can analytically solve Eqs. \eqref{densities1}--\eqref{densities2}. We have the system
\begin{equation}
\label{syst}
    \begin{pmatrix}
    \dot{\rho_{11}}\\\dot{\rho_{22}}\\\dot{\rho_{33}}
\end{pmatrix}= \begin{pmatrix}
-\gamma & \delta & 0\\
\gamma & -\gamma -\delta&\delta \\
0&\gamma&-\delta
\end{pmatrix}\begin{pmatrix}
    \rho_{11}\\\rho_{22}\\\rho_{33}
\end{pmatrix},
\end{equation}
Which can be solved by diagonalising the $3\times3$ matrix. We find its eigenvalues to be $0,-\lambda_{\pm}$ with 
\begin{equation}
\lambda_{\pm}=(\gamma + \delta) \mp \sqrt{\gamma\delta} >0.
\end{equation}
This quantity is the inverse of the relaxation time of our two-spin system. The solutions are

\begin{gather}
\label{analyticdensity}
    \rho_{22}=\rho_{22}^s + Be^{-\lambda_+t}+Ce^{-\lambda_-t},\\
    \rho_{11}=\rho_{11}^s + \frac{\delta B}{\sqrt{\gamma\delta}-\delta}e^{-\lambda_+t}- \frac{\delta C}{\sqrt{\gamma\delta}+\delta}e^{-\lambda_-t},\\
    \rho_{33}=\rho_{33}^s + \frac{\gamma B}{\sqrt{\gamma\delta}-\gamma}e^{-\lambda_+t}- \frac{\gamma C}{\sqrt{\gamma\delta}+\gamma}e^{-\lambda_-t},
\end{gather}
where 
\begin{align}
B&=\frac{\delta-\gamma}{2\sqrt{\gamma\delta}}\left[\left(\rho^s_{22}-\rho_{22}(0)\right)\frac{\delta}{\delta+\sqrt{\delta\gamma}}-\left(\rho_{11}(0)-\rho_{11}^s\right)\right], \\ C&=\frac{\delta-\gamma}{2\sqrt{\gamma\delta}}\left[\left(\rho^s_{22}-\rho_{22}(0)\right)\frac{\delta}{-\delta+\sqrt{\delta\gamma}}+\left(\rho_{11}(0)-\rho_{11}^s\right)\right].
\end{align}
We show the temporal evolution of the populations in Fig. \ref{fig:density} for an initial state with only the upper state of the triplet $\ket{\uparrow\uparrow}$ populated. We refer to the triplet states by their $M_S$ value: Level 1 is $\ket{\uparrow\uparrow}$, Level 0 is $\ket{\uparrow\downarrow^+}$ and Level $-1$ is $\ket{\downarrow\downarrow}$. These results will help us to study the entanglement dynamics.

Moreover, we see that the dynamics at the zero temperature corresponds to an exceptional point \cite{Heiss_2012,Gao_2025,Hatano18082019}. Indeed, the matrix of the differential system \eqref{syst} becomes non-diagonalisable when we take the limit $T \leftarrow 0$, which yields $\delta\leftarrow 0$. The eigenvalues $\lambda_\pm$ collapse into one single eigenvalue $\gamma=\gamma_0^{(2)}$ of doubled multiplicity, however only one eigenvector exists for this eigenvalue. Hence the matrix can only be reduced to its Jordan form  \cite{Heiss_2012}
\begin{equation}
     \begin{pmatrix}
-\gamma & 1 & 0\\
0 & -\gamma &0\\
0&0&0
\end{pmatrix}.
\end{equation}
This indicates a critical regime for the dynamics, where the temporal dependence goes from a sum of exponential to the sum of terms $e^{-\gamma t}$ and $te^{-\gamma t}$. In classical dynamics, this feature is characteristic of a damped oscillator with a quality factor $Q=1/2$, whereas in quantum mechanics it is specific to open systems. Indeed, this exceptional point is allowed only by the the non-unitary evolution of the system. Other cases of Lindbladian dynamics have exhibited an exceptional point too \cite{Gao_2025,Hatano18082019}.

\subsubsection{Dynamics of the off-diagonal coherence}
For the off-diagonal terms, we have
\begin{gather}
\label{corr1}
    \dot{\rho_{13}}=-\frac{\gamma + \delta}{2} \rho_{13}, \hspace{1cm} \dot{\rho_{24}}=-\frac{\gamma + \delta}{2} \rho_{24}, \\
    \dot{\rho_{14}}=-\frac{\gamma}{2} \rho_{14} \hspace{1cm} \dot{\rho_{34}}=-\frac{\delta}{2} \rho_{34}, \\
    \dot{\rho_{12}}= -\left(\gamma+\frac{\delta}{2}\right)\rho_{12} + \delta \rho_{23},\\
    \dot{\rho_{23}}= -\left(\delta+\frac{\gamma}{2}\right)\rho_{23} + \gamma \rho_{12}
    \label{corr2}.
\end{gather}
The consecutive levels of the triplet are coupled to each other in the last two equations, but not the unrelated levels whose off-diagonal coherence vanish exponentially. For the coupled equations, the inverse of the characteristic times are obtained by doing the same procedure as for the diagonal terms and finding the eigenvalues of the matrix system
\begin{equation}
 \begin{pmatrix}
    \dot{\rho_{12}}\\\dot{\rho_{23}}
\end{pmatrix}= \begin{pmatrix}
-\left(\gamma +\frac{\delta}{2}\right) & \delta \\
\gamma & -\left(\delta + \frac{\gamma}{2}\right)
\end{pmatrix}\begin{pmatrix}
    \rho_{12}\\\rho_{23}
\end{pmatrix},   
\end{equation}
which yields the rates
\begin{equation}
\label{rates}
\kappa_{\pm}=\frac{3}{4}(\gamma + \delta) \mp \frac{1}{2}\sqrt{\frac{(\gamma + \delta)^2}{4}+3\gamma\delta}.
\end{equation}
The non-zero eigenvalues of the Liouvillian are given by the rates \eqref{rates} along with ($\gamma + \delta$)/2, $\delta$/2, $\gamma$/2 and $\lambda_\pm$. One can prove that
\begin{equation}
\label{ineq}
\lambda_->\kappa_->\lambda_+>\frac{\gamma + \delta}{2}>\frac{\gamma}{2}>\kappa_+>\frac{\delta}{2}>0.
\end{equation}

 We have thus found the explicit expression of the relaxation and decoherence times for our system, as well as the analytical expression of the population densities. We observe that the off-diagonal coherence decrease slower than the populations, which can be part of the explanation for the next observations. 
 
 Therefore, the spectral gap, defined as the eigenvalue corresponding to the largest finite characteristic time, is
 \begin{equation}
 \label{gap}
     \frac{\delta}{2}= \frac{\gamma_0^{(2)} N_0}{2}, 
 \end{equation}
at finite temperature, but it is switched to $\gamma_0^{(2)}/2$ at zero temperature. We note that the strict zero temperature condition gives a higher spectral gap than at an infinitesimal temperature. Indeed, while the temperature is above zero, the spectral gap is the one defined in Eq. \eqref{gap}. When the temperature reaches zero, the eigenvalue generating this gap degenerates to the zero eigenvalue, which means that we have to take the next non-zero eigenvalue, which is $\gamma/2=\gamma_0^{(2)}/2$. 

\subsubsection{Concurrence}

We quantify the bipartite entanglement between the two spins with the concurrence $C$. We first return to the Schrödinger picture with the unitary transformation $\rho_s(t)=e^{-\frac{i}{\hbar}H_St}\rho e^{\frac{i}{\hbar}H_St} $. We hereafter drop the subscript $s$ for brevity and work with the density matrix in the Schrödinger picture. By letting $\sigma_y$ denote the $y$ component of the Pauli matrix, we define $\tilde{\rho}:=(\sigma_y \otimes \sigma_y) \rho^* (\sigma_y \otimes \sigma_y)$. We have to be careful while doing the tensor product of $\sigma_y$ because we are working with the eigenbasis of $H_S$. In this basis we have
$$\sigma_y \otimes \sigma_y=
\begin{pmatrix}
0 & 0 & -1&0\\
0 & 1&0&0 \\
-1&0&0&0\\
0&0&0&-1
\end{pmatrix}.
$$
We then let $\{\xi_1,\xi_2,\xi_3,\xi_4\}$ the eigenvalues of $\rho\tilde{\rho}$ in the descending order. The Wootter's concurrence is then given by \cite{wootters_entanglement_2001}
\begin{equation}
C=\max \left[0,\sqrt{\xi_1}-\sqrt{\xi_2}-\sqrt{\xi_3}-\sqrt{\xi_4}\right].
\end{equation}

For the stationary state, $\rho$ is diagonal in the eigenbasis. We show that the concurrence for the stationary state is
\begin{equation}
      C^s=\begin{cases}\rho_{44}-3\rho_{22}^s, \textrm{ if }\rho_{44}>\frac{N_0(1+N_0)}{\frac{1}{3}+2N_0(1+N_0)},\\
    0, \textrm{ otherwise.}
\end{cases}
\end{equation}  
We see that if $\rho_{44}=1$ then $C^s=1$, which is expected because only the singlet state, which is maximally entangled, is populated. Moreover, for high $N_0$ we have $C^s=2\rho_{44}-1$, which means that in order to have a non-zero concurrence we have to initially populate the singlet state at least to 1/2. For $N_0\ll1$, $C^s=\rho_{44}-3N_0(1-\rho_{44}) + o(N_0)$. To summarise, the behaviour of the concurrence for the stationary state is only determined by the initial population of the singlet, which is expected because this entangled level is not interacting with the others, and therefore it will not be destroyed by the thermal decoherence.

This shows that it is possible to precisely control and reach a stable degree of entanglement between the two systems only by controlling the initial population of the singlet. This result is in accordance with Ref. \cite{li_stationary_2008}, where a further study of the steady states is performed, and with Ref. \cite{an_entanglement_2007}, where a study of two qutrit at $T=0$ K is done.

So far, we have not specified which state was the ground state. Thanks to the Heisenberg interaction between the qubits, one can turn the decoherence-free singlet state into the ground state of the system by tuning the magnetic field $B_0$ to fulfill the condition $\hbar \omega_0 < J$.

\subsection{Numerical results for the concurrence}
In the present subsection, we show the temporal evolution of the concurrence for different initial states. We observe the phenomenon of delay, sudden death and revival of the concurrence. We analyse it based on the Werner states. We choose $J=0.1\hbar$ and we will use different values of $N_0$.
\subsubsection{Four examples of initial states}
Let us first give four examples of initial states, each of which reveals an aspect of the concurrence.  
\begin{itemize}
 \item Let $\ket{\chi}=\frac{1}{2}\left(\ket{\uparrow\uparrow}+\ket{\downarrow\downarrow}-\sqrt{2}\ket{\uparrow\downarrow^-}\right)$ a linear combination of Bell states and 
 \begin{equation}
 \rho(0)=\ket{\chi}\bra{\chi} = \begin{pmatrix}
\frac14 & 0 & \frac14 & -\frac{\sqrt2}{4}\\
0 & 0 & 0 & 0\\
\frac14 & 0 & \frac14 & -\frac{\sqrt2}{4}\\
-\frac{\sqrt2}{4} & 0 & -\frac{\sqrt2}{4} & \frac12
\end{pmatrix}, 
\end{equation}
for which the singlet state is half populated. For small values of $N_0$ and long time, the concurrence reaches the population of the singlet. With $N_0$ increasing, the concurrence tends to stay at its minimum value (Fig. \ref{conc4}). The oscillations are due to the interaction between the qubits.
  \begin{figure}[htbp]
        \centering
        \includegraphics[width=\linewidth]{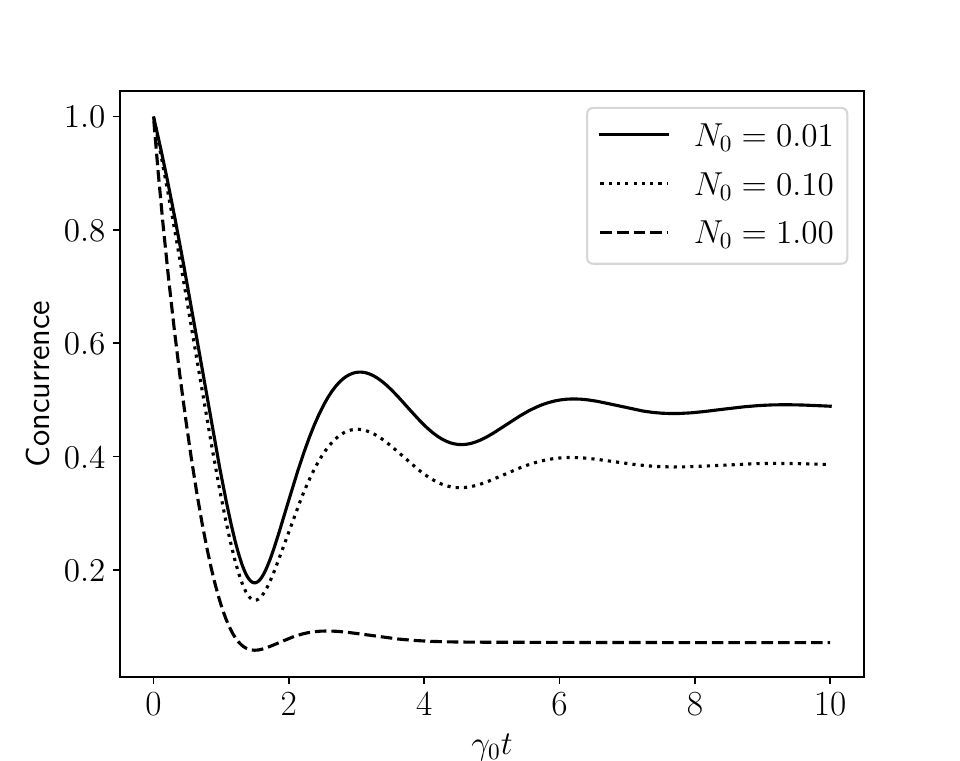}
        \caption{Time dependence of the concurrence for the initial state $\rho(0)=\ket{\chi}\bra{\chi}$. }
        \label{conc4}
    \end{figure}

\item Let $\ket{\phi}=\frac{1}{\sqrt2}\left(\ket{\uparrow}+\ket{\downarrow}\right), \ket{\phi\phi}=\ket{\phi}\otimes\ket{\phi}$ and
\begin{equation}
\label{ini2}
\rho(0)=\ket{\phi\phi}\bra{\phi\phi}=\begin{pmatrix}
\frac14 & \frac{\sqrt2}{4} & \frac14 & 0\\
\frac{\sqrt2}{4} & \frac12 & \frac{\sqrt2}{4} & 0\\
\frac14 & \frac{\sqrt2}{4} & \frac14 & 0\\
0 & 0 & 0 & 0
\end{pmatrix}.
\end{equation}
Because the singlet state is not originally populated, the concurrence strictly vanishes at the stationary state (Fig. \ref{initial1} (a)). Moreover the concurrence does not oscillate and reaches a maximum due to the temporary occupation of the entangled state $\ket{\uparrow\downarrow^+}$. The off-diagonal coherence $\rho_{13}$ between Level 0 and Level $-1$ reaches a maximum just before the concurrence does (Fig. \ref{initial1} (b)).
    \begin{figure}[htbp]
        \centering
        \begin{subfigure}{0.47\textwidth}
            \centering
            \includegraphics[width=\textwidth]{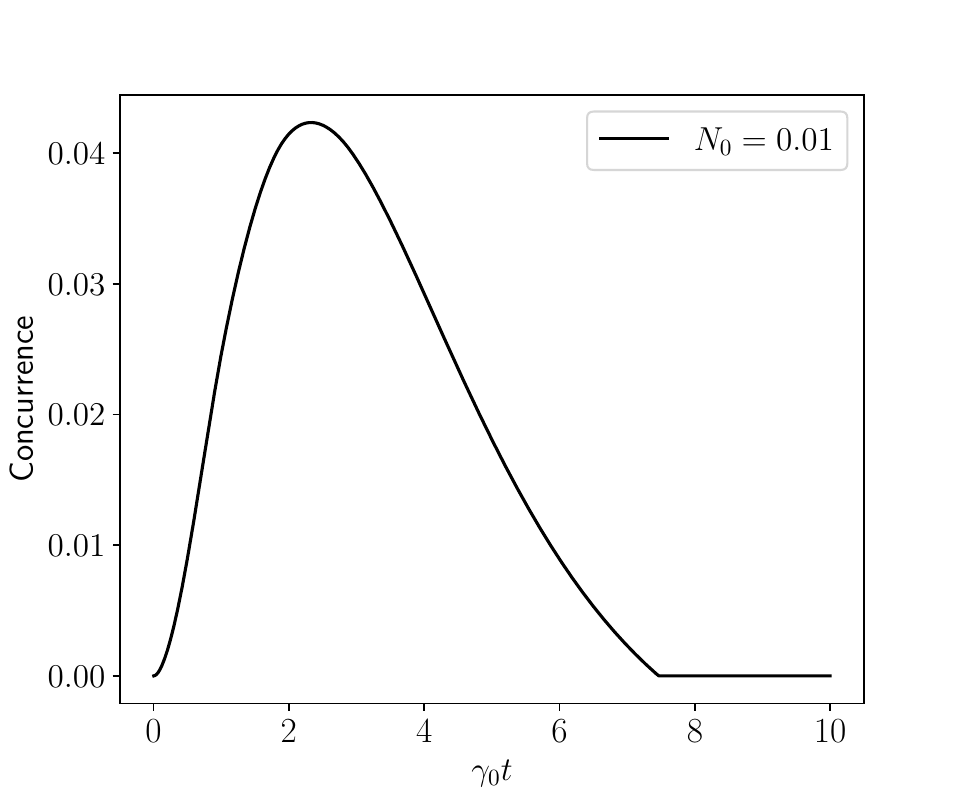}
            \caption{}
        \end{subfigure}
        \hfill
        \begin{subfigure}{0.47\textwidth}
            \centering
            \includegraphics[width=\textwidth]{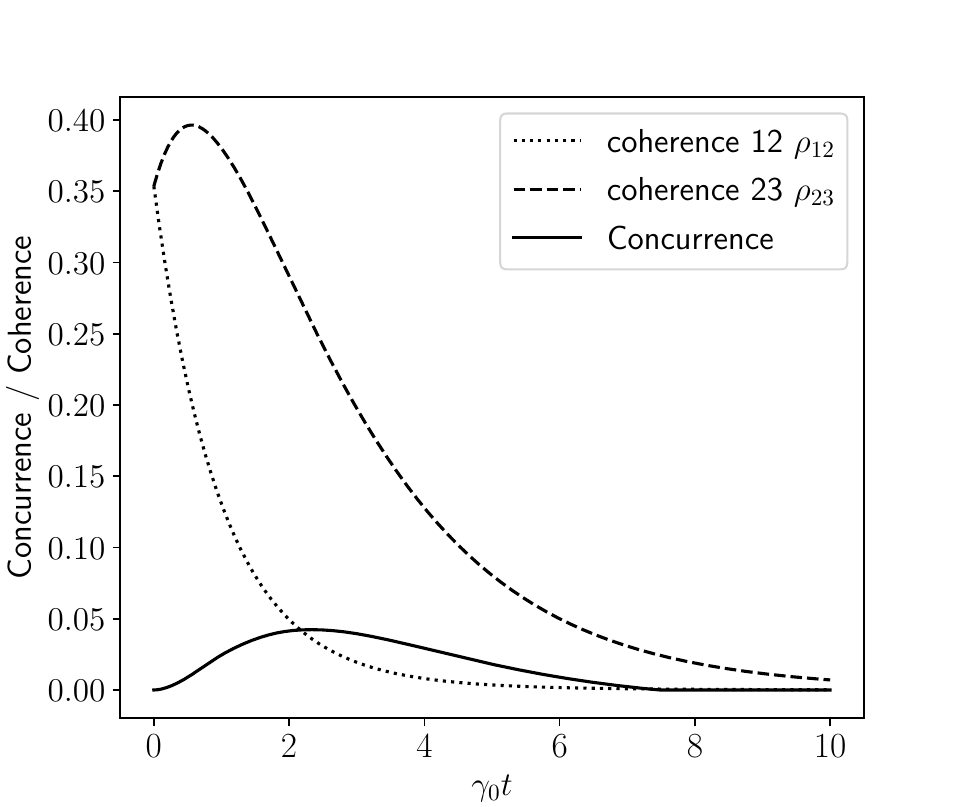}
            \caption{}
        \end{subfigure}
        \caption{ (a) Time dependence of the concurrence for the initial state $\rho(0)=\ket{\phi\phi}\bra{\phi\phi}$. (b) Off-diagonal coherence between levels 1-0 and 0-($-1$) with $N_0=0.01$.}
        \label{initial1}
    \end{figure}

\item Let $\ket{\psi}=\frac{1}{\sqrt2}\left(\ket{\uparrow}-\ket{\downarrow}\right)$ and
\begin{equation}
\rho(0)=\ket{\phi\psi}\bra{\phi\psi}=\begin{pmatrix}
\frac14 & 0 & -\frac14 & -\frac{\sqrt2}{4}\\
0 & 0 & 0 & 0\\
-\frac14 & 0 & \frac14 & \frac{\sqrt2}{4}\\
-\frac{\sqrt2}{4} & 0 & \frac{\sqrt2}{4} & \frac12
\end{pmatrix}.
\end{equation}
The singlet state is half populated but this time the concurrence is zero at the start and increases (Fig. \ref{conc2}). Like before, the increase of $N_0$ suppresses the oscillation and lowers the stationary-state value.  
    \begin{figure}[htbp]
        \centering
        \includegraphics[width=\linewidth]{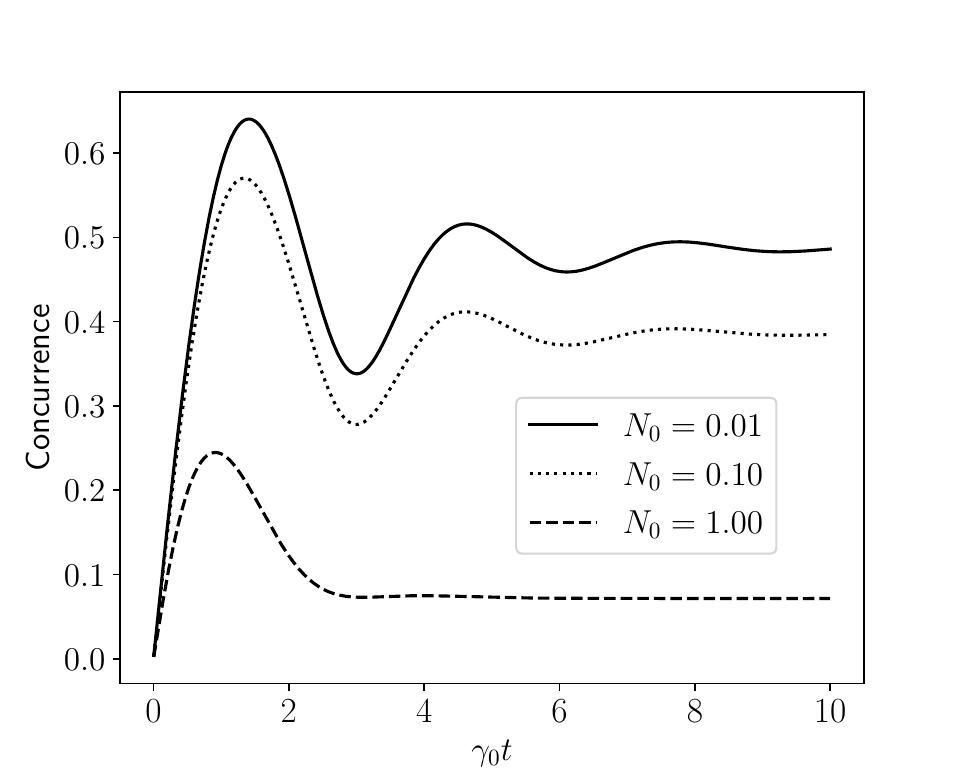}
        \caption{Time dependence of the concurrence for the initial state $\rho(0)=\ket{\phi\psi}\bra{\phi\psi}$. }
        \label{conc2}
    \end{figure}

\item Let $\rho(0)=\frac{1}{2}\ket{\phi\phi}\bra{\phi\phi}+\frac{1}{2}\ket{\chi}\bra{\chi}$.
\begin{equation}
\label{mixed}
\rho(0)=\begin{pmatrix}
\frac14 & \frac{\sqrt2}{8} & \frac14 & -\frac{\sqrt2}{8}\\
\frac{\sqrt2}{8} & \frac14 & \frac{\sqrt2}{8} & 0\\
\frac14 & \frac{\sqrt2}{8} & \frac14 & -\frac{\sqrt2}{8}\\
-\frac{\sqrt2}{8} & 0 & -\frac{\sqrt2}{8} & \frac14
\end{pmatrix}.
\end{equation}
Each level is initially equally populated. For low $N_0$, the concurrence drops to zero and then rises, namely sudden death and revival, while it stays at zero once $N_0$ reaches a large enough value (Fig. \ref{initial3}). 
\begin{figure}[htbp]
    \centering
    \includegraphics[width=\linewidth]{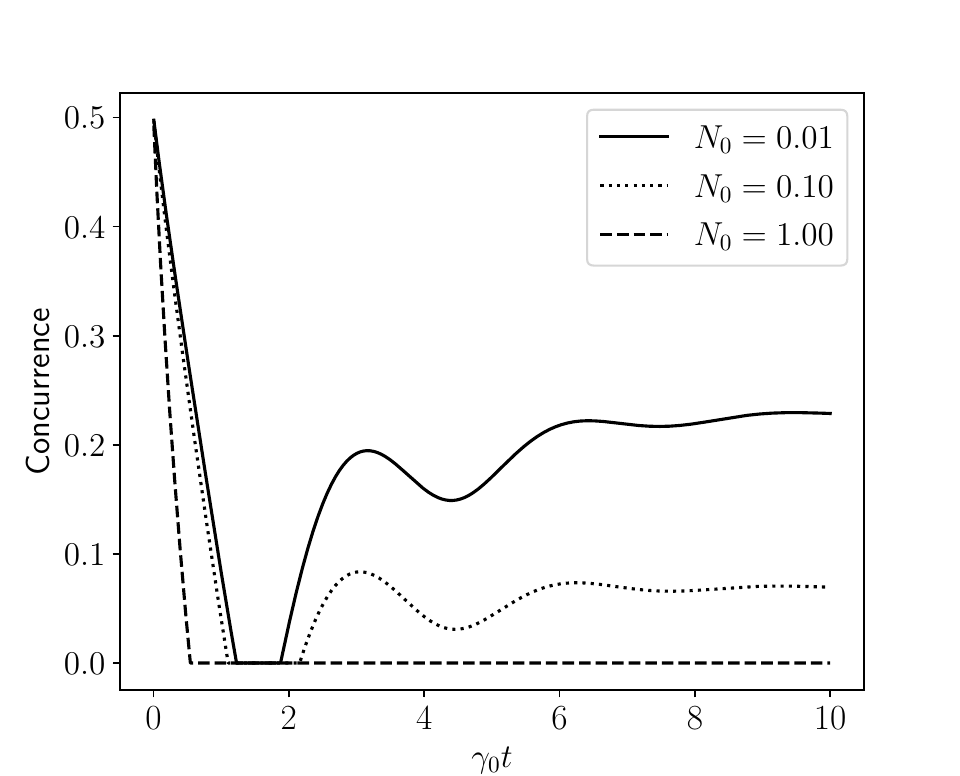}
    \caption{Time dependence of the concurrence for the initial mixed state $\rho(0)=\frac{1}{2}\ket{\phi\phi}\bra{\phi\phi}+\frac{1}{2}\ket{\chi}\bra{\chi}$.}
    \label{initial3}
\end{figure}
\end{itemize}

We see the major role of the singlet. Aside from determining the equilibrium value, it influences the dynamics through the off-diagonal coherence, even without directly exchanging its population with the other levels. If not initially populated, as the second initial state Eq. \eqref{ini2}, the concurrence still temporarily reaches non-zero values due to the entangled state of the triplet. We also see that the initial preparation influences the monotony of the concurrence, and even totally suppresses it during a certain time (Fig. \ref{initial3}). 

We stress here that our model shows the sudden death and revival of the entanglement in the level of the weak coupling Born-Markov approximation \eqref{gksl}. Yu and Eberly predicted the sudden death of the concurrence for two non-interacting qubits in two different cavities following Markovian dynamics \cite{Yu_2004}, while the phenomenon of revival has been numerically observed in Ref.  \cite{PhysRevA.79.042302} and experimentally observed in Ref. \cite{PhysRevLett.104.100502} for the same non-interacting systems without the Born-Markov approximation. Moreover, in the case of interacting qubits the non-Markovian study conducted in Ref. \cite{wu_exact_2013} provides the same kind of behaviour, with a difference that the concurrence, as expected, reaches a non-zero stationary value if the singlet state is populated. For the non-Markovian case, the revival can be explained as a memory effect of the reservoirs. 

For the non-interacting Markovian case, Ref. \cite{PhysRevA.79.042302} shows that the interaction of the decoupled qubits with a common reservoir creates an effective coupling between them, responsible for the revival of the concurrence, without oscillations. This previous study has been made using the exact pseudomode method \cite{PhysRevA.55.2290}. Moreover, the sudden death and revival of the entanglement was demonstrated for a two-qubit system following the Lehmberg–Agarwal master equation in a vacuum field \cite{PhysRevA.74.024304}. Our results show that the perturbative approach is enough to accurately describe the dynamics of the spins and of the entanglement. We also show that, in this model too, the Heisenberg interaction causes the concurrence to oscillate.

\subsubsection{Werner states}
Another class of initial states that we will study are the Werner states \cite{PhysRevA.40.4277}: 
\begin{equation}
\rho(0)=W_\pm=r\ket{\uparrow\downarrow^\pm}\bra{{\uparrow\downarrow^\pm}}+\frac{1-r}{4}I\otimes I.
\end{equation}
We present different numerical results, quantitatively explain them and derive some analytical properties for the concurrence at zero temperature by using the results for the dynamics (see \ref{dynamics}). In particular, we find the exact condition for the concurrence to be delayed, to disappear and to appear, and we use it to find the time of these events.

The Werner states are diagonal and hence the off-diagonal terms stay at zero and the concurrence does not oscillate. Still, we observe the sudden death and revival, meaning that it is not only due to the coherence between the eigenstates of the Hamiltonian. The density matrix stays diagonal at all times, and therefore the eigenvalues of $\rho \tilde{\rho}$ are $(\rho_{44})^2$, $(\rho_{22})^2$ and $\rho_{11}\rho_{33}$ with a multiplicity of two. 

We will use different values of $-1/3<r<1$ and $N_0$. The initial concurrence is $C_\pm(0) = \max \left[0,\frac{3r-1}{2}\right]$.
\begin{figure}[htbp]
    \centering
    \begin{subfigure}{0.47\textwidth}
        \centering
        \includegraphics[width=\textwidth]{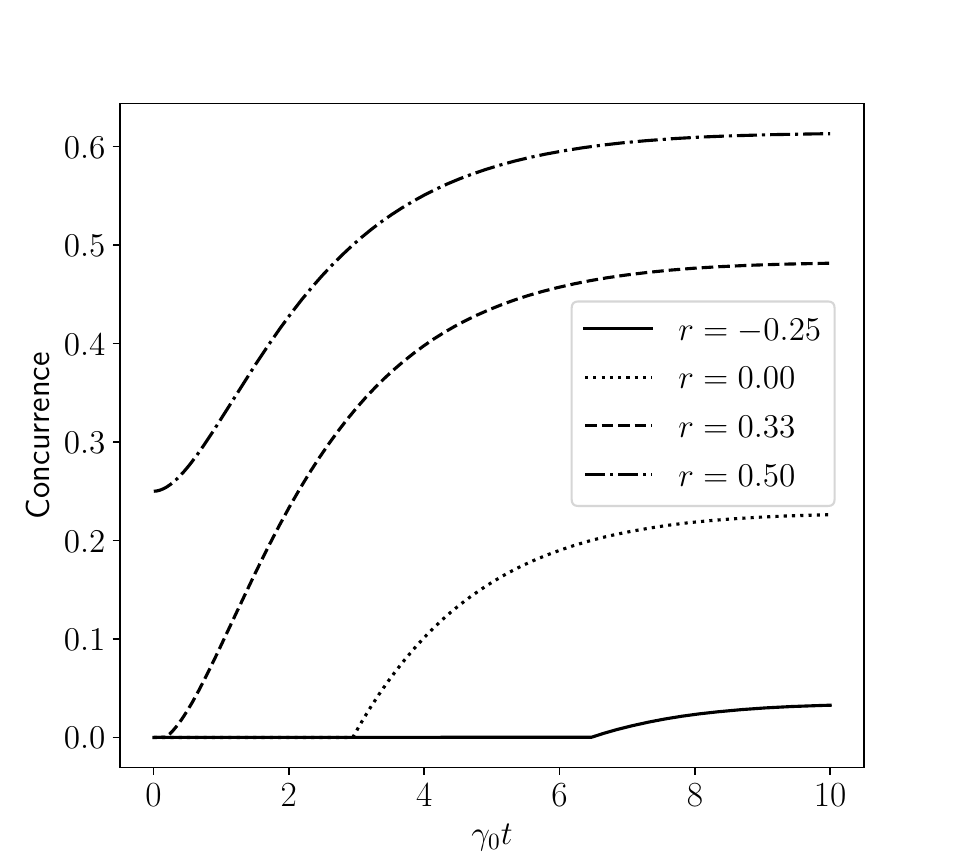}
        \caption{}
    \end{subfigure}
    \hfill
    \begin{subfigure}{0.47\textwidth}
        \centering
        \includegraphics[width=\textwidth]{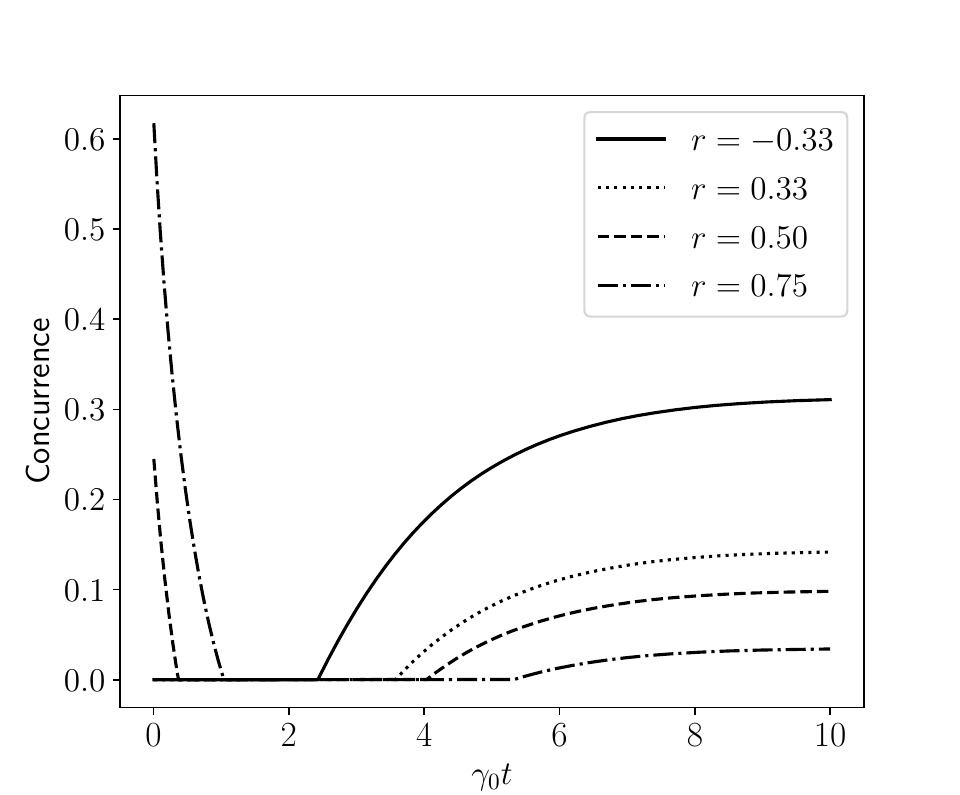}
        \caption{}
    \end{subfigure}
    \caption{Time dependence of the concurrence for the initial state (a) $W_-$ and (b) $W_+$ with $N_0$=0.01.}
    \label{concwerner}
\end{figure}

For $W_-$, the concurrence is delayed for $r<1/3$ (Fig. \ref{concwerner} (a)). For $W_+$, it starts from a finite value for $r>1/3$, falls to zero and then rises (Fig. \ref{concwerner} (b)). We see the same behaviour of sudden death and revival as for the initially mixed state from Eq. \eqref{mixed} in Fig. \ref{initial3}. 

\begin{figure}[htbp]
        \centering
        \begin{subfigure}{0.47\textwidth}
            \centering
            \includegraphics[width=\textwidth]{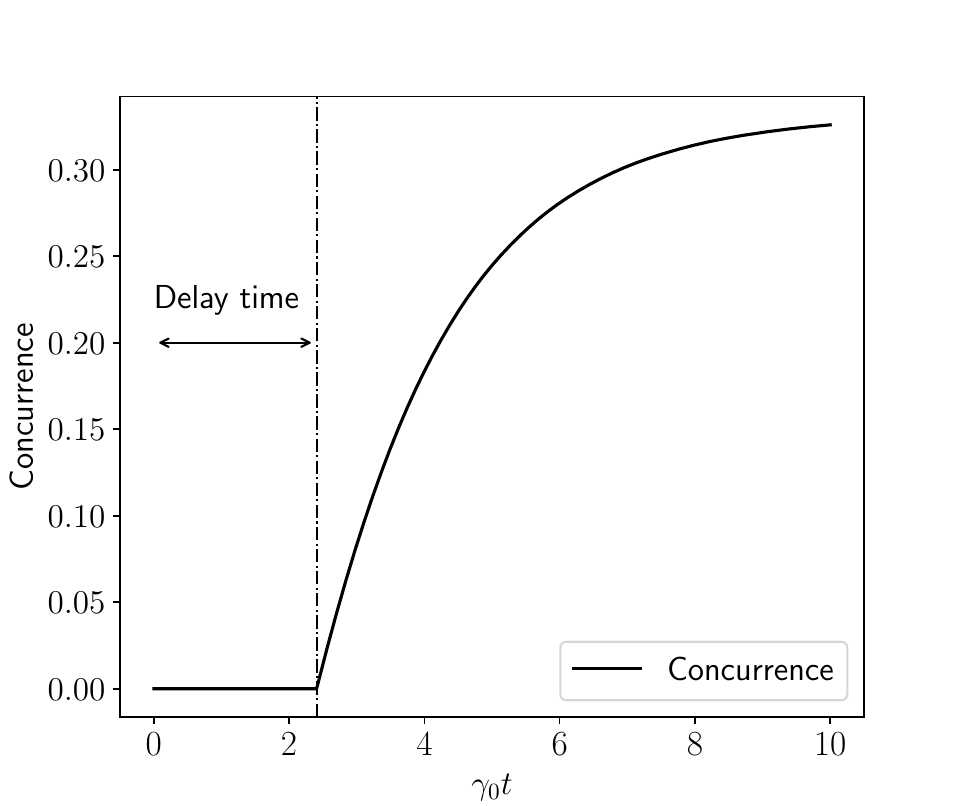}
            \caption{}
        \end{subfigure}
        \hfill
        \begin{subfigure}{0.47\textwidth}
            \centering
            \includegraphics[width=\textwidth]{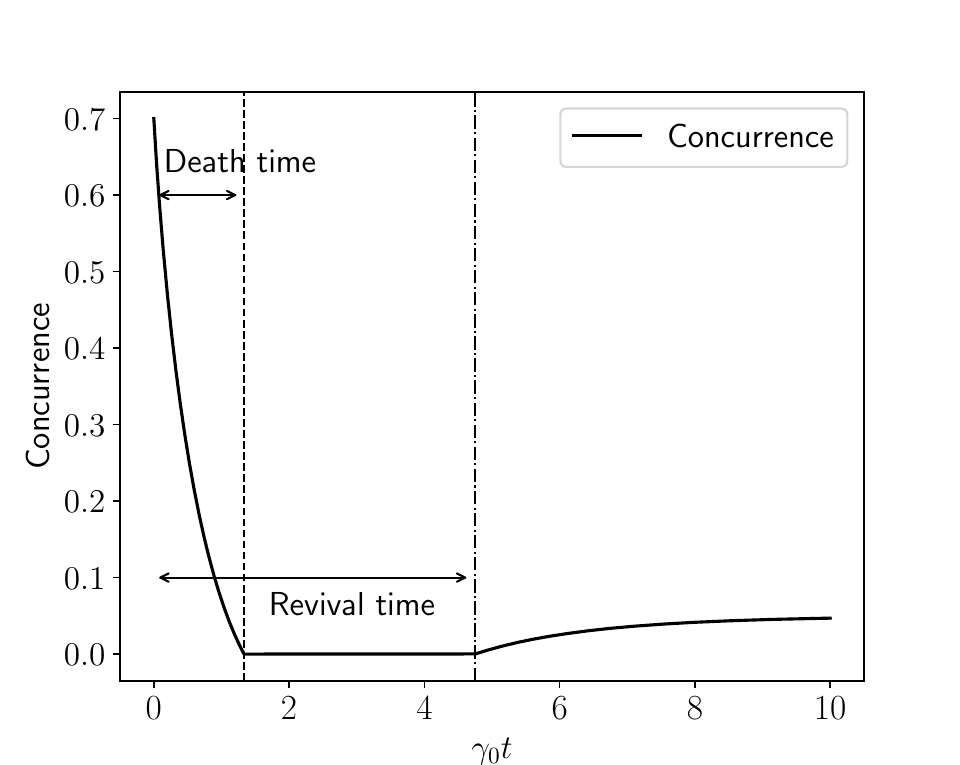}
            \caption{}
        \end{subfigure}
        \caption{ (a) Schematic illustration of the delay time. (b) Schematic illustration of the death time and the revival time.}
        \label{def}
    \end{figure}

As exemplified in Fig. \ref{def}, we define the delay time as the time at which the concurrence takes a non-zero value for the first time. The death time is the time at which the concurrence reaches zero, while the revival time is the time at which the concurrence takes a non-zero value after a death occurred before. These times cannot be analytically expressed even for $N_0=0$. 

\begin{figure}[htbp]
    \centering
    \begin{subfigure}{0.47\textwidth}
        \centering
        \includegraphics[width=\textwidth]{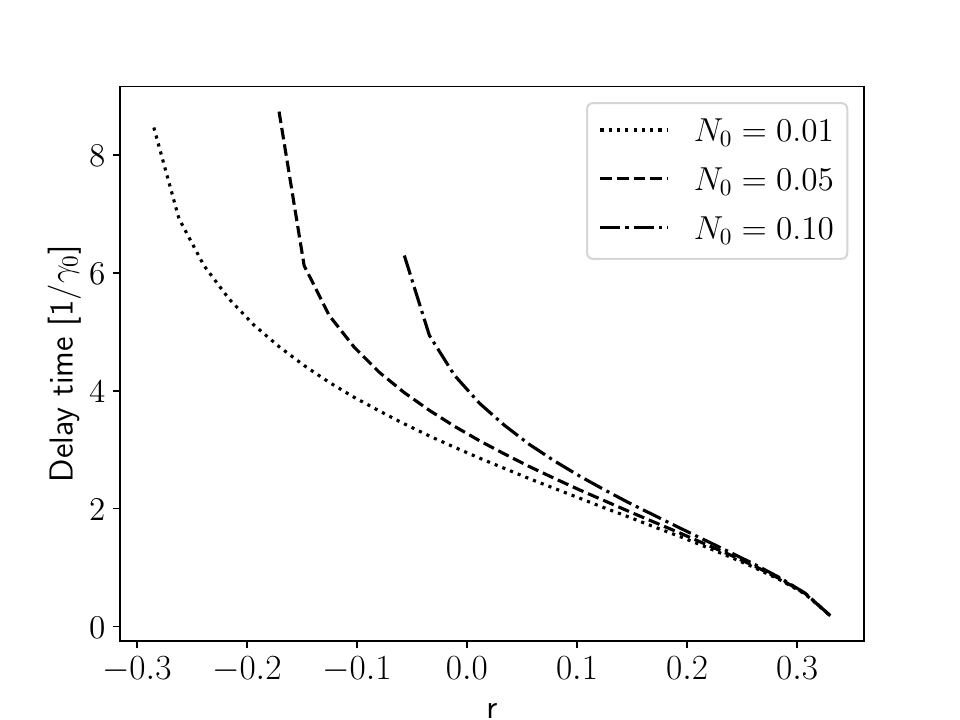}
        \caption{}
    \end{subfigure}
    \hfill
    \begin{subfigure}{0.47\textwidth}
        \centering
        \includegraphics[width=\textwidth]{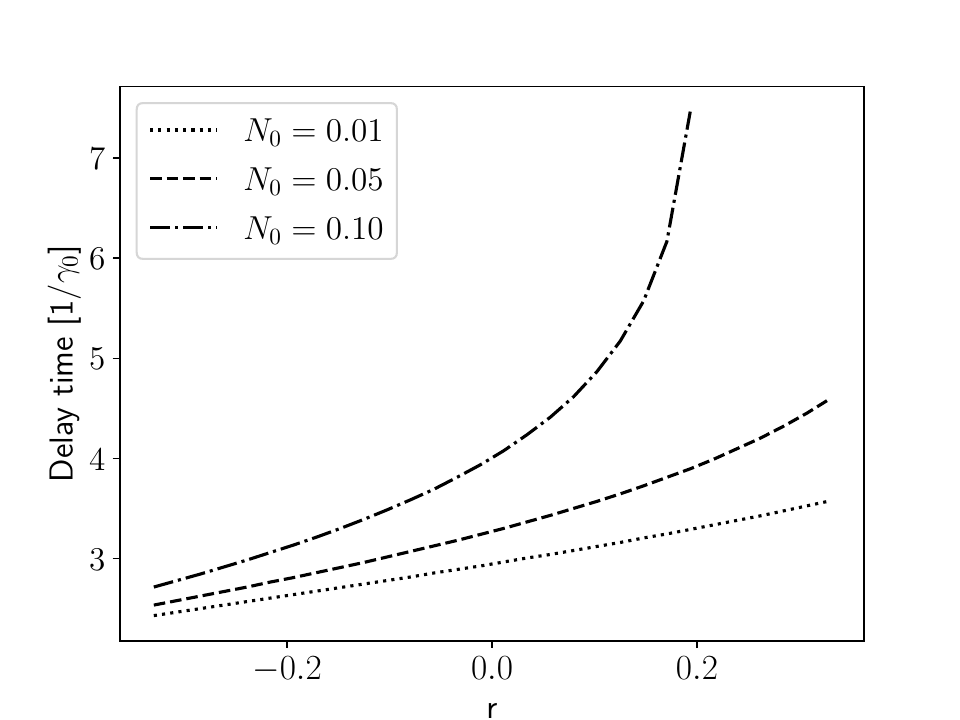}
        \caption{}
    \end{subfigure}
    \caption{Delay time of the concurrence for the initial state (a) $W_-$ and (b) $W_+$.}
    \label{werner12delay}
\end{figure}

For $W_-$ (Fig. \ref{werner12delay} (a)), the delay time diverges as $r$ approaches $-1/3$ because this limit means that the singlet state is not populated, so that the concurrence stays at zero. The increase in $N$ seems to accelerate the divergence. On the other hand, the limit $r=1/3$ means that the singlet state is half populated, which forces the concurrence to instantly rise. For $W_+$ (Fig. \ref{werner12delay} (b)) and small $N_0$, the delay time seems linear in $r$. 

\begin{figure}[htbp]
    \centering
    \begin{subfigure}{0.47\textwidth}
        \centering
        \includegraphics[width=\textwidth]{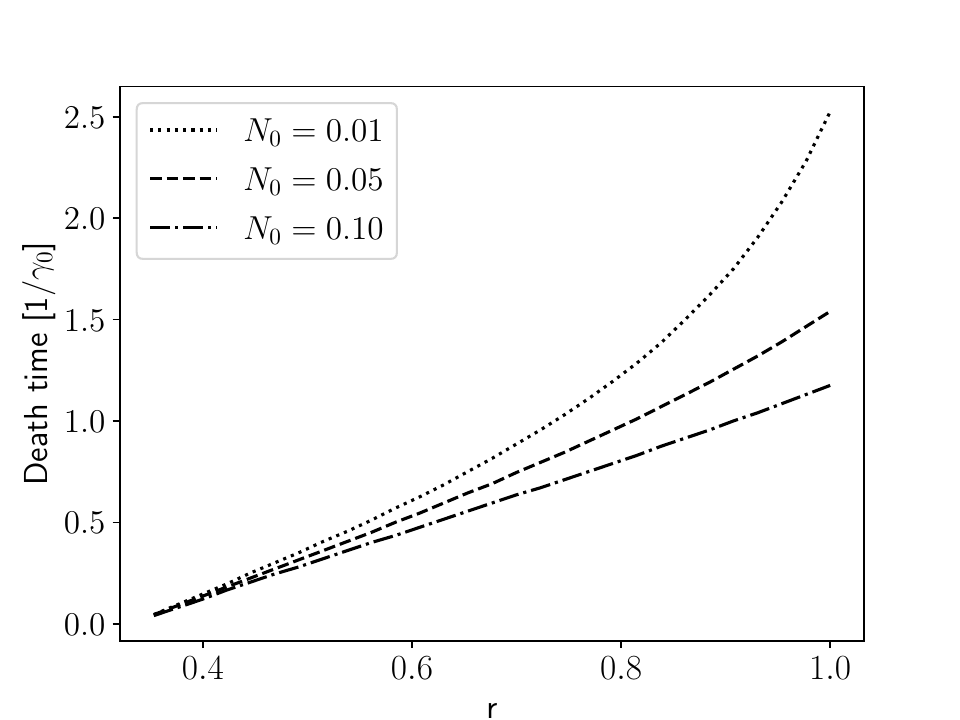}
        \caption{}
    \end{subfigure}
    \hfill
    \begin{subfigure}{0.47\textwidth}
        \centering
        \includegraphics[width=\textwidth]{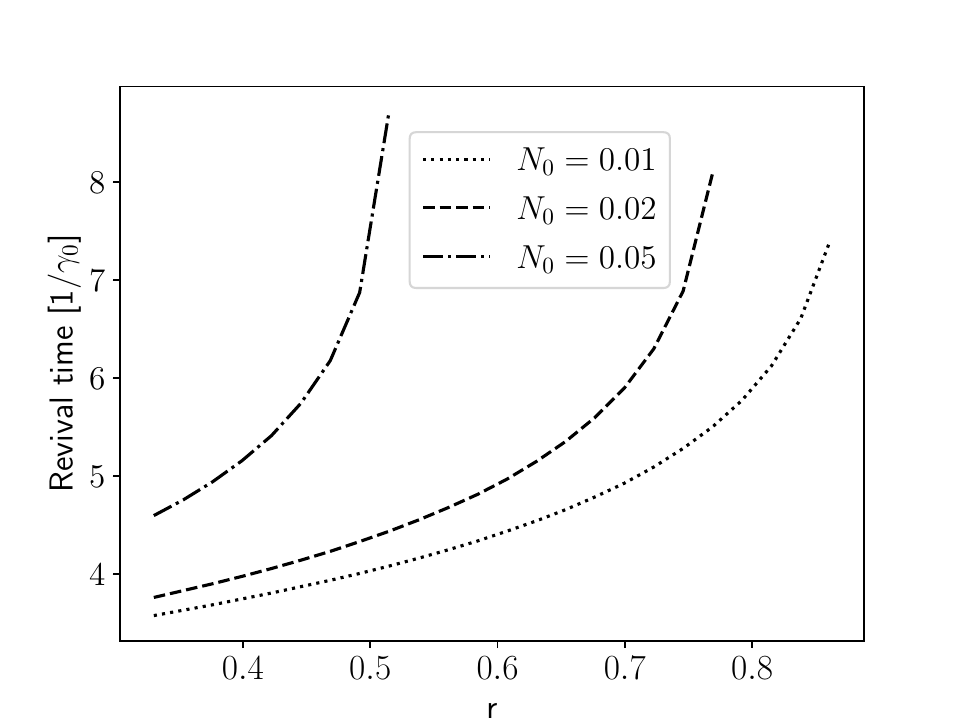}
        \caption{}
    \end{subfigure}
    \caption{(a) Death time and (b) revival time of the concurrence for $W_+$.}
    \label{werner2death}
\end{figure}

The death time of the concurrence occurring for $r>1/3$ presents a linear growth in $r$ for high $N_0$ as shown in Fig. \ref{werner2death} (a). We note that the death time for $r=1$ is finite for non-zero values of $N_0$ but diverges as $N_0$ approaches zero, which means that the concurrence vanishes in a infinite time in vacuum when only the entangled state of the triplet is initially populated. The revival time, however, diverges as $r$ tends to $1$ for all values of $N_0$. The increase of $N_0$ also dramatically accelerates the divergence. 

\begin{figure}[htbp]
    \centering
    \includegraphics[width=\linewidth]{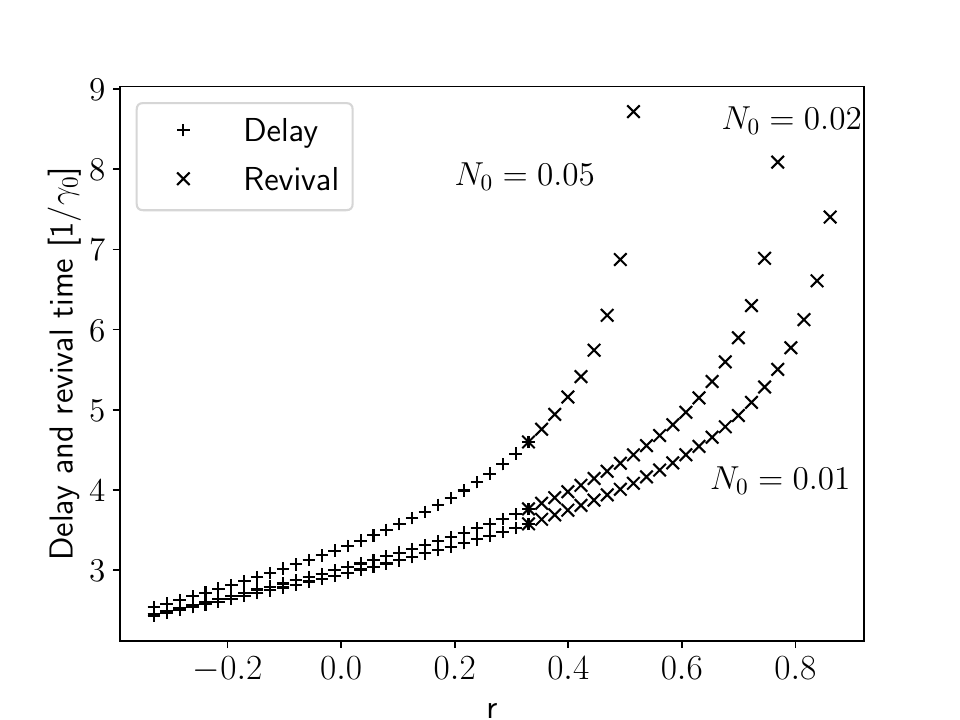}
    \caption{Delay time (in the range $r<1/3$) and revival time (in the range $r>1/3$) of the concurrence for $W_+$.}
    \label{combinedWerner1}
\end{figure}

By plotting together the delay time for $r<1/3$ (Fig. \ref{werner12delay} (b)) and the revival time for $r>1/3$ (Fig. \ref{werner2death} (b)), we find a continuous growth with the divergence appearing as $r$ tends to unity (Fig. \ref{combinedWerner1}). This suggests that the relation between $r$ and the delay time and the one between $r$ and the revival time are the same.    

For the Werner states, the explanation of the sudden death and revival resides in the ordering between  $(\rho_{44})^2$, $(\rho_{22})^2$ and $\rho_{11}\rho_{33}$. There are two scenarios in which the concurrence is non-zero:
    \begin{equation}
    \label{1scenar}
    \rho_{44}>\rho_{22}+2\sqrt{\rho_{11}\rho_{33}},
    \end{equation}
and    
    \begin{equation}
    \label{2scenar}
    \rho_{22}>\rho_{44}+2\sqrt{\rho_{11}\rho_{33}}.
    \end{equation}

Physically, this shows that a non-zero concurrence happens when one of the two entangled states is sufficiently populated, while the condition $\sqrt{\rho_{11}\rho_{33}}>\max[\rho_{22},\rho_{44}]$ is enough to ensure $C=0$, meaning that the geometric mean of the populations of the classical states $\ket{\uparrow\uparrow}$ and $\ket{\downarrow\downarrow}$ defines the condition of extinction. This explains the delay for $W_-$ and the death and revival for $W_+$ in Fig. \ref{concwerner}. 

\begin{figure}[htbp]
    \centering
    \begin{subfigure}{0.47\textwidth}
        \centering
        \includegraphics[width=\textwidth]{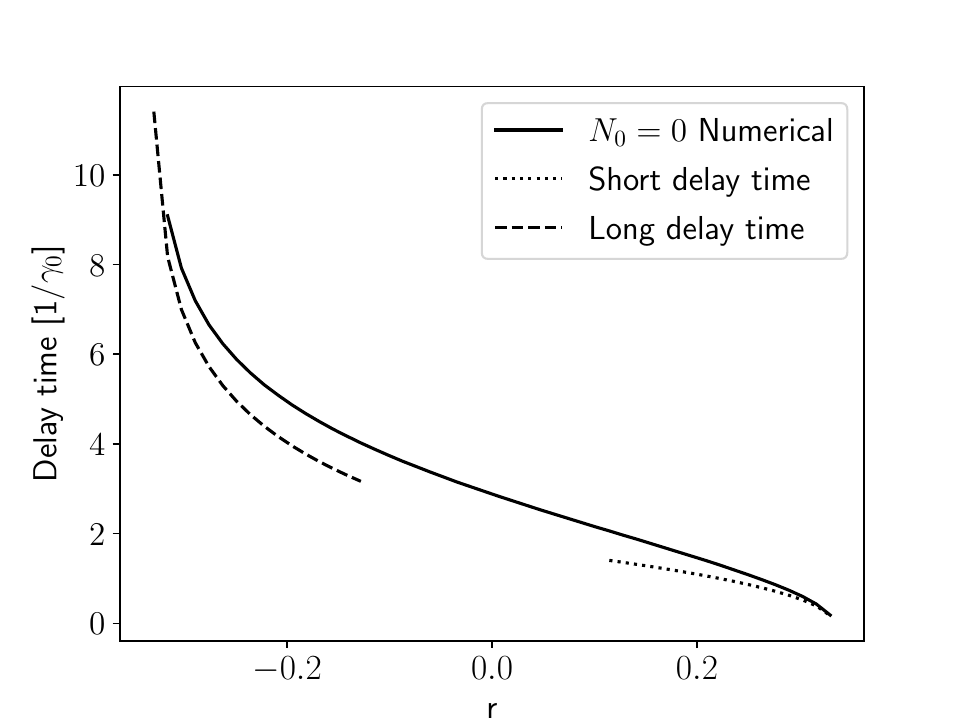}
        \caption{}
    \end{subfigure}
    \hfill
    \begin{subfigure}{0.47\textwidth}
        \centering
        \includegraphics[width=\textwidth]{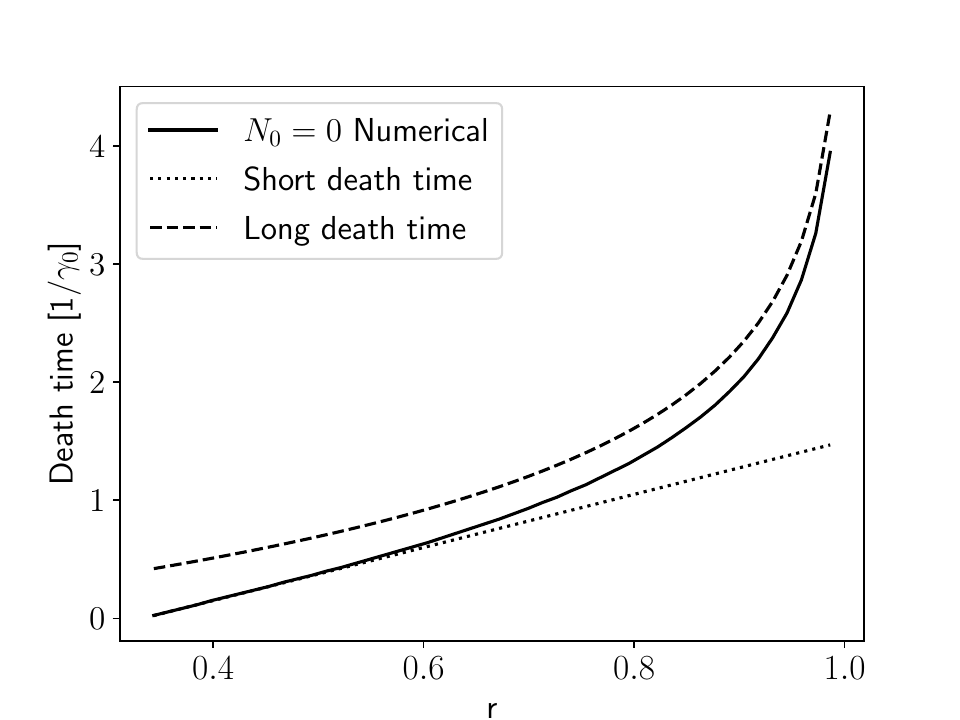}
        \caption{}
    \end{subfigure}
    \caption{(a) Delay time for $W_-$ and (b) death time for $W_+$ at zero temperature.}
    \label{dev}
\end{figure}

For $W_-$, one can see that for $r>1/3$ the singlet state is populated at $\rho_{44}>1/2$, more than all the other states combined. With the populations $\rho_{11}$ and $\rho_{22}$ decreasing, the condition of the first scenario \eqref{1scenar} is true for all times, as observed in Fig. \ref{concwerner}. For $r<1/3$, the concurrence starts at zero and the increase of $r$ increases the population of the singlet state, while the other populations tend to decrease, thus fulfilling the first condition in a finite time, which we defined as the delay time. For $N_0=0$, by using Eq. \eqref{analyticdensity}, we can show that the delay time $t_d$ satisfies
\begin{equation}
    \frac{1+3r}{1-r}=e^{-\gamma_0t_d}(1+\gamma_0t_d+2\sqrt{3e^{\gamma_0t_d}-2-\gamma_0t_d}).
\end{equation}
For short delay time and by defining $\tau_d:=\gamma_0 t_d$, we perform an asymptotic analysis, which leads to 
\begin{equation}
    \tau_d=\sqrt{3(1-3r)},
\end{equation}
while for the long delay time we have
\begin{equation}
    \tau_d=-2 \ln{\left(\frac{1}{3}+r\right)}.
\end{equation}
We show in Fig. \ref{dev} the numerical results and the asymptotic analysis.

For $W_+$, the sudden death and revival are explained by the second scenario \eqref{2scenar}. Firstly for $r<1/3$, the delay time is explained by the same process as before. However for $r>1/3$, initially the entangled state of the triplet is more than half populated and therefore the second scenario \eqref{2scenar} rules. Its population then depletes toward the lower state, thus breaking the inequality and causing the sudden death. Finally, the revival occurs in the same condition as before: the first scenario \eqref{1scenar} rules when $N_0$ is small enough. This explains also the fact that the delay time and the revival time follow the same dependence on $r$ as observed in Fig. \ref{combinedWerner1}. Same as before, for $N_0=0$ we obtain the asymptotic expansion of the dimensionless death time $\tau_{death}$ for short time
\begin{equation}
    \tau_{death}= \frac{3}{4}(3r-1),
\end{equation}
and for long time
\begin{equation}
    \tau_{death}= -\ln{(1-r)}.
\end{equation}
These results are illustrated in Fig. \ref{dev}, which confirms that the death time linearly scales with $r$ when $r\approx 1/3$.

\subsection{XXZ interaction}
It is interesting to consider the XXZ spin interaction, which may model a more realistic electronic dipole-dipole interaction \cite{muller_quantum_nodate}. Instead of Eq. \eqref{xxx}, we introduce:
\begin{equation}
H_S=\hbar\omega_0(S_1^z+S_2^z) + J_{xy}(S^x_1S^x_2+S^y_1S^y_2) + J_zS^z_1S^z_2,
\end{equation}
which is diagonal in the triplet-singlet eigenbasis, with the eigenvalues 
\begin{equation}
   \left\{\hbar\omega_0+\frac{J_z}{4},\frac{2J_{xy}-J_z}{4},-\hbar\omega_0+\frac{J_z}{4},-\frac{2J_{xy}+J_z}{4}\right\}.  
\end{equation}

The consequence of this new coupling is that the transition frequencies between the three levels of the triplet become different. Instead of having one frequency $\omega_0$ we now have
\begin{equation}
    \omega_1=\omega_0+\frac{J_z-J_{xy}}{2\hbar}\textrm{, }\omega_2=\omega_0-\frac{J_z-J_{xy}}{2\hbar}.
\end{equation}
 This modifies the master equation because we now have to account for the slowly oscillating terms at $\omega_2 - \omega_1$.  We still apply the rotating-wave approximation to terms oscillating at $\omega_2 + \omega_1$, $2\omega_{1}$ and $2\omega_{2}$. The selection rules still hold and hence we have the two jump operators:
\begin{align}
  \vec{A}(\omega_1)&=\frac{g\mu_B}{\sqrt2}(\hat{x}+i\hat{y})\ket{\uparrow\downarrow^+}\bra{\uparrow\uparrow},\\ 
  \vec{A}(\omega_2)&=\frac{g\mu_B}{\sqrt2}(\hat{x}+i\hat{y})\ket{\downarrow\downarrow}\bra{\uparrow\downarrow^+}.
\end{align}

Before going further, we can see that we have two regimes. If $\omega_1$ and $\omega_2$ are sufficiently separated, we can also apply the rotating-wave approximation to the terms oscillating at $\omega_2 - \omega_1$, which means that we consider two decoupled two-level systems (Levels 1-0 and Levels 0-($-1$)). This approximation can simplify the dynamics but is only relevant when $\abs{J_{xy}-J_z}$ is significantly higher than $\hbar$ multiplied by the inverse of a relaxation time. This is not true in general and in order to stay close to the precedent case, we consider that both frequencies are near $\omega_0$, which means a small perturbation from the  coupling. 

Instead of applying the rotating-wave approximation to the $e^{i(\omega_2 - \omega_1)t}$ terms we take $e^{\pm i(\omega_2 - \omega_1)t}\approx1$. We then start with
\begin{equation}
\begin{aligned}
    \dot{\rho_S}&=\sum_{\omega,\omega'}\sum_{i,j}e^{i(\omega'-\omega)t}\Gamma_{ij}(\omega)
    \\&\left(S_j(\omega)\rho_SS_i^\dagger(\omega')-S_i^\dagger(\omega')S_j(\omega)\rho_S\right) + \text{h.c.}
\end{aligned}
\end{equation}
and $\Gamma_{ij}$ defined in Eq. \eqref{cor tensor}, which reduces to
\begin{equation}
\begin{aligned}
    \dot{\rho_S}&=\sum_{\omega,\omega'}e^{i(\omega'-\omega)t}\Gamma(\omega)\left(\vec{S}(\omega)\rho_S\vec{S}^\dagger(\omega')-\vec{S}^\dagger(\omega')\vec{S}(\omega)\rho_S\right) \\&+ \text{h.c.}
\end{aligned}
\end{equation}
with
\begin{equation}
\begin{aligned}
   \Gamma(\omega)&=\frac{1}{6\pi^2 \epsilon_0 \hbar c^3}\frac{1}{c^2}
\int_0^\infty d\omega_k \omega_k^3
\bigg[(1+N(\omega_k))
\\
&\left.
\int_0^\infty ds\, e^{-i(\omega_{k}-\omega)s}
+N(\omega_k) \int_0^\infty ds\,
e^{i(\omega_k+\omega)s}
\right].
\end{aligned}
\end{equation}
By writing for $i=1,2$
\begin{gather}
\gamma_{i}=\gamma_0^{(2)}(1+N(\omega_{i})),\\
\delta_{i}=\gamma_0^{(2)}N(\omega_{i}),
\end{gather}
and $N(\omega_i)=N_i$, this leads the diagonal terms to
\begin{gather}
\label{densz}
    \dot{\rho_{11}}=\delta_1\rho_{22} - \gamma_1\rho_{11},\hspace{1cm} \dot{\rho_{33}}=\gamma_2\rho_{22} - \delta_2\rho_{33}, \\
    \dot{\rho_{22}}=\gamma_1\rho_{11}-\gamma_2\rho_{22} + \delta_2\rho_{33}-\delta_1\rho_{22}, \\ \label{densz2} \dot{\rho_{44}}=0,
\end{gather}
and the off-diagonal terms to
\begin{gather}
    \dot{\rho_{13}}=-\frac{\gamma_1 + \delta_2}{2} \rho_{13}, \hspace{1cm} \dot{\rho_{24}}=-\frac{\gamma_2 + \delta_1}{2} \rho_{24}, \\
    \dot{\rho_{14}}=-\frac{\gamma_1}{2} \rho_{14}, \hspace{1cm} \dot{\rho_{34}}=-\frac{\delta_2}{2} \rho_{34}, \\
    \dot{\rho_{12}}= -\frac{\gamma_1+\gamma_2+\delta_1}{2}\rho_{12} + \frac{\delta_1+\delta_2}{2} \rho_{23},\\
    \dot{\rho_{23}}= -\frac{\gamma_2+\delta_1 +\delta_2}{2}\rho_{23} + \frac{\gamma_1+\gamma_2}{2} \rho_{12}.
\end{gather}
By taking the limit $J_{xy}=J_z$ ($\omega_1=\omega_2$) we retrieve the precedent case Eqs. \eqref{densities1}--\eqref{densities2} and Eqs. \eqref{corr1}--\eqref{corr2}.

On the other hand, by considering the decoupled regime where the rotating-wave approximation is applicable to the oscillations at $\omega_2-\omega_1$, one can find that the diagonal terms behave the same as in Eqs. \eqref{densz}--\eqref{densz2} but the evolution of $\rho_{12}$ and $\rho_{23}$ are not coupled anymore, which can be explained by the fact that the states of the triplet are too decoupled to influence each other. 

In any case, the off-diagonal terms still vanish for the stationary state, and the stationary populations are given by 
\begin{gather}
    \rho^s_{22}=\frac{1-\rho_{44}}{1+\frac{\delta_1}{\gamma_1}+\frac{\gamma_2}{\delta_2}}=\frac{N_2(1+N_1)}{1+2N_2+N_1+3N_2N_1}(1-\rho_{44}),\\
    \rho^s_{11}= \frac{\delta_1}{\gamma_1}\rho^s_{22}=\frac{N_2N_1}{1+2N_2+N_1+3N_2N_1}(1-\rho_{44}),\\
    \rho^s_{33}= \frac{\gamma_2}{\delta_2}\rho^s_{22}=\frac{(1+N_1)(1+N_2)}{1+2N_2+N_1+3N_2N_1}(1-\rho_{44}).
\end{gather}
The stationary concurrence becomes more complicated to describe. We have the three scenarios:
\begin{equation}
  \sqrt{\rho^s_{11}\rho^s_{33}}> \max[\rho^s_{22}, \rho_{44}] \implies C^s=0,  
\end{equation}
 \begin{equation}
 \label{xxyscenar}
 \begin{aligned}
 \rho^s_{22}&> \max[\sqrt{\rho^s_{11}\rho^s_{33}}, \rho_{44}] \\
\implies C^s&=\max \left[0, \rho^s_{22}-\rho_{44}-2\sqrt{\rho^s_{11}\rho^s_{33}}\right],
\end{aligned}
 \end{equation}
 \begin{equation}
     \begin{aligned}
      \rho_{44}&> \max[\sqrt{\rho^s_{11}\rho^s_{33}}, \rho^s_{22}] \\
\implies C^s&=\max \left[0, \rho_{44}-\rho^s_{22}-2\sqrt{\rho^s_{11}\rho^s_{33}}\right].
     \end{aligned}
 \end{equation}
 
The condition $\rho^s_{22}> \sqrt{\rho^s_{11}\rho^s_{33}}$ is equivalent to $N_2>N_1$. One can see that, contrary to the previous system, it is possible to reach a non-zero stationary concurrence without populating the singlet. Indeed, the second scenario \eqref{xxyscenar} shows that the entangled state of the triplet can determine the concurrence. For instance, by putting $\rho_{44}=0$, we see that for $N_1<1/3$ and $N_2>4N_1/(1-3N_1)$, the concurrence is finite. This result shows that for an XXZ coupling, it is possible to reach a stationary entanglement between the two qubits without even populating the singlet. We can perform a numerical study and shows that the temporal evolution of the concurrence exhibits the same behaviour as before, with sudden death and revival or delay. 

\section{Conclusions and outlook}
In this study, we derived the spectral correlation tensor and the spontaneous emission rate for the Zeeman coupling to a quantised magnetic field. We then used this result to derive the Markovian dynamics of two interacting two-level systems in the XXX and XXZ couplings. We found analytical expressions of the eigenvalues of the Liouvillian, which are the spin relaxation times, the decoherence times and spectral gap. We also numerically showed that the entanglement between the qubits presents various behaviours for different initial conditions. In particular, the perturbative Born-Markov approximation approach is enough to observe a sudden death and revival of the entanglement, while the SU(2) coupling causes the concurrence to oscillate and allows us to have a decoherence-free ground state. The benefit of this approach is that it provided analytical results for this model without losing the behaviour found by the exact methods. In the present approach, we quantitatively studied the dependence of the delay, death and revival of the entanglement regarding the Werner parameter $r$.

It would be interesting to study the dynamics for many interacting two-level systems. Qualitatively, one can say that the Zeeman coupling $\vec{S}\cdot\vec{B}$ ensures the validity of the selection rules and therefore only allows step-by-step transitions inside one spin multiplet. Then, it would be reasonable to write the GKSL jump operators as the ladder operators of each spin multiplet. Each multiplet then has its own GKSL dissipator, hence making the population of each level of a multiplet only depends on the other populations of the same multiplet. The singlet state would play the same role as in this study by determining the stationary entanglement and the conditions of delay, death and revival of the concurrence.

\section*{Acknowledgements}
We would like to thank Dvira Segal for reading the manuscript and kindly advicing us.

\bibliography{bibliography}
\end{document}